\documentclass[twocolumn]{aastex701}
\usepackage{graphicx}
\usepackage{placeins}
\usepackage{amsmath}
\usepackage{tgcursor}
\usepackage{enumitem}

\begin{document}

\title{JWST Edge-on Disk Ice (JEDIce): Program overview and ice survey results} 

\author[0000-0002-8716-0482]{Jennifer B.~Bergner}
\affiliation{University of California, Berkeley, Berkeley CA 94720, USA}
\email{jbergner@berkeley.edu}

\author[0000-0003-2631-5265]{Nicole Arulanantham}
\affiliation{Astrophysics \& Space Institute, Schmidt Sciences, New York, NY 10011, USA}
\email{narulanantham@schmidtsciences.org}

\author[0000-0003-1197-7143]{Emmanuel Dartois}
\affiliation{Institut des Sciences Mol\'{e}culaires d’Orsay, CNRS, Univ. Paris-Saclay, 91405 Orsay, France}
\email{emmanuel.dartois@universite-paris-saclay.fr}

\author[0000-0001-7479-4948]{Maria N. Drozdovskaya}
\affiliation{Physikalisch-Meteorologisches Observatorium Davos und Weltstrahlungszentrum (PMOD/WRC), Dorfstrasse 33, 7260 Davos Dorf,
Switzerland}
\email{maria.drozdovskaya.space@gmail.com}

\author[0000-0001-6307-4195]{Daniel Harsono}
\affiliation{Institute of Astronomy, Department of Physics, National Tsing Hua University, Hsinchu, Taiwan}
\email{dharsono@gapp.nthu.edu.tw}

\author[0000-0003-1878-327X]{Melissa McClure}
\affiliation{Leiden Observatory, Leiden University, PO Box 9513, 2300 RA Leiden, The Netherlands}
\email{mcclure@strw.leidenuniv.nl}

\author[0000-0003-4985-8254]{Jennifer A. Noble}
\affil{Physique des Interactions Ioniques et Mol\'{e}culaires, CNRS, Aix-Marseille Univ., 13397 Marseille, France}
\email{jennifer.noble@univ-amu.fr}

\author[0000-0001-8798-1347]{Karin I. \"Oberg}
\affiliation{Center for Astrophysics, Harvard \& Smithsonian, 60 Garden St., Cambridge, MA 02138, USA}
\email{koberg@cfa.harvard.edu}

\author[0000-0001-7552-1562]{Klaus M. Pontoppidan}
\affiliation{Jet Propulsion Laboratory, California Institute of Technology, 4800 Oak Grove Drive, Pasadena, CA 91109, USA}
\email{klaus.m.pontoppidan@jpl.nasa.gov}

\author[0000-0001-8227-2816]{Yao-Lun Yang}
\affiliation{Star and Planet Formation Laboratory, RIKEN Pioneering Research Institute, Wako-shi, Saitama, 351-0106, Japan}
\email{yaolunyang.astro@gmail.com}

\author[0000-0002-2131-4346]{Korash Assani}
\affiliation{Astronomy Department, University of Virginia, Charlottesville, VA 22904, USA}
\affiliation{Virginia Institute of Theoretical Astronomy, University of Virginia, Charlottesville, VA 22904, USA}
\email{ka8km@virginia.edu}

\author[0000-0002-7402-6487]{Zhi-Yun Li}
\affiliation{Astronomy Department, University of Virginia, Charlottesville, VA 22904, USA}
\affiliation{Virginia Institute of Theoretical Astronomy, University of Virginia, Charlottesville, VA 22904, USA}
\email{zl4h@virginia.edu}

\author[0000-0002-3401-5660]{Julia C.~Santos}
\altaffiliation{51 Pegasi b Fellow}
\affiliation{Center for Astrophysics, Harvard \& Smithsonian, 60 Garden St., Cambridge, MA 02138, USA}
\email{julia.santos@cfa.harvard.edu}

\author[0000-0002-0786-7307]{Will E. Thompson}
\affiliation{University of California, Berkeley, Berkeley CA 94720, USA}
\email{willet@berkeley.edu}

\author[0009-0008-8810-6577]{Lukas Welzel}
\affiliation{Leiden Observatory, Leiden University, PO Box 9513, 2300 RA Leiden, The Netherlands}
\email{welzel@strw.leidenuniv.nl}

\author[0009-0008-4937-3314]{Elizabeth S. Yunerman}
\affiliation{Center for Astrophysics, Harvard \& Smithsonian, 60 Garden St., Cambridge, MA 02138, USA}
\email{eyunerman@cfa.harvard.edu}

\author[0000-0001-8407-4020]{Aditya M. Arabhavi}
\affiliation{Jet Propulsion Laboratory, California Institute of Technology, 4800 Oak Grove Drive, Pasadena, CA 91109, USA}
\email{aditya.mahadeva.arabhavi@jpl.nasa.gov}

\author[0000-0003-2014-2121]{Alice S. Booth} 
\affiliation{Center for Astrophysics \textbar\, Harvard \& Smithsonian, 60 Garden St., Cambridge, MA 02138, USA}
\email{alice.booth@cfa.harvard.edu}

\author[0000-0002-8023-2834]{Charles Mentzer}
\affiliation{Institute of Astronomy, Department of Physics, National Tsing Hua University, Hsinchu, Taiwan}
\email{cmentzer@gapp.nthu.edu.tw}

\author[0000-0002-0554-1151]{Mayank Narang}
\affiliation{Jet Propulsion Laboratory, California Institute of Technology, 4800 Oak Grove Drive, Pasadena, CA 91109, USA}
\email{mayank.narang@jpl.nasa.gov}

\author[0000-0002-1493-300X]{Thomas Henning}
\affiliation{Max Planck Institut für Astronomie, Heidelberg, Germany}
\email{henning@mpia.de}

\author[0000-0001-7455-5349]{Inga Kamp}
\affiliation{Kapteyn Astronomical Institute, Rijksuniversiteit Groningen, Groningen, The Netherlands}
\email{kamp@astro.rug.nl}

\author[0000-0002-8545-6175]{Giulia Perotti}
\affiliation{Max Planck Institut für Astronomie, Heidelberg, Germany}
\altaffiliation{Niels Bohr Institute, University of Copenhagen, Copenhagen,Denmark}
\email{perotti@mpia.de}

\author[0000-0003-2090-2928]{Alice Somigliana}
\affiliation{Max Planck Institut für Astronomie, Heidelberg, Germany}
\email{alsomigliana@mpia.de}

\begin{abstract}
\noindent The icy material within protoplanetary disks plays a central role in planet formation, yet remains poorly characterized by observations.  We present 1.6--28~$\mu$m spectra of five disks obtained as part of the JWST Edge-on Disk Ice (JEDIce) program, representing the largest survey of disk ices to date.  The major ice species H$_2$O, CO$_2$, and CO are detected towards all disks, and exhibit a wide range of absolute optical depths and optical depth ratios across the sample.  This is suggestive of a range of ice abundances and compositions, but quantitative constraints will require radiative transfer modeling.  All disks exhibit ice features across the entire spatial region where the IR continuum is detected; vertically elevated ice grains therefore seem to be ubiquitous in disks.  The CO ice is consistently dominated by apolar CO:CO$_2$ mixtures, implying that the disk ice compositions are neither completely reset nor pristinely inherited from the protostellar stage.  The presence of these mixtures also suggests that entrapment may be important in shaping the spatial distribution of CO within the disks.  Small molecules commonly seen in protostellar ices (CH$_4$, CH$_3$OH, NH$_3$) are generally not detected in our sample, though tracers of ammonium salts (OCN$^-$ and the 6.85 $\mu$m band) are common, potentially reflecting an evolution towards comet-like ice compositions.  The spectra also contain a wealth of information about the micron-sized dust, atomic and molecular gas, and PAH content, which together with the ice constraints will provide a comprehensive picture of the chemical, physical, and dynamical state of these systems.
\end{abstract}

\section{Introduction}
\label{sec:intro}
The materials present within protoplanetary disks are the building blocks from which planets assemble.  Throughout most of the disk, the high densities and low temperatures cause the majority of volatiles to remain frozen as icy mantles coating the refractory dust grains \citep{Aikawa1999}.  These ices play a wide-ranging role in shaping the outcomes of planet formation: the freeze-out of volatiles onto dust grains increases the total mass of solids available to form planetary cores, and potentially influences the sticking efficiency during the early stages of grain growth \citep[e.g.][]{Stevenson1988,Wada2013,Ros2013,Gundlach2015,Pinilla2017}.  Moreover, as the major reservoir of volatiles, the abundance, distribution, and speciation of disk ice impacts the elemental budget (e.g.~C, N, O, H) of planet-forming solids and gas \citep{Oberg2011,Bitsch2022,Pacetti2025}, as well as the inventory of complex molecules incorporated into icy planetesimals \citep{Rubin2019,Krijt2023}.  Constraints on protoplanetary disk ices are therefore critical to predicting and interpreting both the physical and chemical outcomes of planet formation.

The vibrational modes of ice-phase molecules can be observed in absorption at near- and mid-infrared wavelengths.  For disks, this means that a close to edge-on viewing geometry is needed such that the solid icy material is eclipsing the host star.  This viewing geometry also produces intrinsically low fluxes; only a few detections of ice absorption features in disks were possible with previous IR telescopes, and these were limited by a lack of sensitivity and spectral resolution \citep[e.g.][]{Terada2007,Aikawa2012, Terada2017}.  The complex propagation of IR radiation through edge-on disks further complicates the interpretation of their ice absorption features \citep{Pontoppidan2005}.  Fortunately, there has been significant foundational work in modeling edge-on disk observations at optical/near-IR wavelengths \citep{Stapelfeldt1998,Wolf2003,Luhman2007,Pontoppidan2007,Duchene2010}, which has informed recent approaches to modeling observed ice absorption features \citep[e.g.][]{Dartois2022,Arabhavi2022,Sturm2023b}.  In a few disks, ice features have also been detected at far-IR wavelengths \citep{Malfait1999,McClure2012,McClure2015,Min2016}, though extension to a larger sample is not currently possible given the lack of a far-IR facility.   

In the past few years, the James Webb Space Telescope (JWST) has transformed our ability to characterize ice features in edge-on disks, thanks to its exceptional sensitivity, angular resolution, spectral resolution, and spectral bandwidth.  \citet{Sturm2023c,Sturm2024} reported the first near-/mid-IR JWST observations of an edge-on disk, HH 48 NE.  In addition to the abundant ice species H$_2$O, CO$_2$, and CO, low-abundance species were also detected for the first time in a disk, including $^{13}$CO$_2$, OCN$^-$, OCS, and NH$_3$.  Moreover, the detection of the H$_2$O, CO$_2$, and CO ice bands at high signal-to-noise and spectral resolution enabled detailed modeling of their band profiles to assess the ice mixing status \citep{Bergner2024}.  Detections of H$_2$O, CO$_2$, and CO ice features with NIRSpec have since been reported towards the Class II disks FS Tau B, HH 30, Tau 042021, and d216-0939 \citep{Pascucci2025, Dartois2025,Potapov2025}, and the 3~$\mu$m H$_2$O feature was also detected towards the 114–426 disk with NIRCam imaging \citep{Ballering2025}.

From these studies, several important questions are emerging.  First, spatial mapping has revealed ice features throughout the entire vertical extent where the IR continuum is detected \citep{Sturm2023c,Dartois2025}.  This is surprising since ices are not expected to survive in the highly irradiated disk upper layers, and implies either some source of shielding or continuous replenishment of icy grains from the midplane \citep[e.g.][]{Bergner2021, Woitke2022}.  Even more unexpected is that, in the same studies, CO ice features are detected throughout almost the entire region where H$_2$O and CO$_2$ are detected, despite its much higher volatility.  One compelling possibility is that CO ice is trapped within the H$_2$O and/or CO$_2$ matrix and survives in the solid state above temperatures where it would normally sublimate, a phenomenon predicted by lab experiments \citep[e.g.][]{Collings2004,Simon2019}. Indeed, there is spectroscopic evidence for CO entrapment in CO$_2$ and H$_2$O ice in the HH 48 NE disk \citep{Bergner2024}.  Still, it is clear that we do not yet understand how even the most abundant ices are distributed in disks, and the factors (grain dynamics, ice mixing/entrapment, chemical evolution) that influence their distributions.  Another key question is whether the minor ice species reported towards HH 48 NE are common in other disks, and whether species that are expected to be present at higher abundances (NH$_3$, CH$_3$OH) are seen in other disks.   Beyond inventorying the composition of planetary building blocks, this is critical to understanding the ice-phase chemical evolution between protostars and planetesimals.  Answering these questions and generalizing our understanding to the broader disk population requires high-quality observations of a larger sample of edge-on disks.

In this paper, we introduce the `JWST Edge-on Disk Ice' (JEDIce) program, a survey of six disks representing approximately half of the known population of edge-on protoplanetary disks.  The primary motivation for this program is to understand the range of possible ice compositions in protoplanetary disks, and how the ice inventory and spatial distribution are related to other disk properties.  The disks were targeted in JWST's Cycle 4 with deep NIRSpec and MIRI spectroscopic observations, providing comprehensive constraints on the ice compositions and the broader physical/chemical context.  

An overview of the JEDIce program including observational design and execution is provided in Section \ref{sec:obs}.  Section \ref{sec:res} presents a first look at the complete observational data set: we show that the near- to mid-IR spectra contain numerous ice absorption features, as well as a wealth of additional constraints on the small dust population, molecular gas in the inner disk, tracers of outflowing jets or winds, and PAHs.  In Section \ref{sec:ice}, we carry out an empirical analysis of the H$_2$O, CO$_2$, and CO ice band profiles and optical depth maps, focusing on implications for the mixing status and spatial distribution of ices in the disks.  We additionally search for absorption features from low-abundance ice species.  Section \ref{sec:concl} summarizes our key findings and future avenues.   

\section{Observations}
\label{sec:obs}
\subsection{Disk targets}
The program was designed to provide insight into the demographics of ice compositions across different planet-forming systems.  To this end, a sample of disk targets was selected based on the criteria of (i) edge-on viewing inclinations ($>$80$^\circ$) determined from previous optical, near-infrared, and millimeter imaging, (ii) disk major axes $>1''$ to enable spatially resolved analyses, (iii) no or minimal remnant protostellar envelope, and (iv) low-mass stars (M-K type) representative of the most common planet hosts. This resulted in 6 sources spanning K4-M1 spectral types and hosted in 4 different star-forming regions.  Details about the disk targets are provided in Table \ref{tab:disk_sample}.

\begin{deluxetable*}{llccccc}
	\tablecaption{JEDIce disk sample. \label{tab:disk_sample}}
	\tablecolumns{7} 
	\tablewidth{\textwidth} 
	\tablehead{
        \colhead{Name}        & 
        \colhead{Coordinates (J2000)} & 
        \colhead{Region}      &
        \colhead{Inc.~($^\circ$)} &
        \colhead{Dist.~(pc)} &
        \colhead{Sp.~Type} &
        \colhead{References}
        }
\startdata
HV Tau C           & 04:38:35.51, +26:10:41.32 & Taurus      &  84  & 140 & M0 & 1,7 \\
Oph 163131         & 16:31:31.25, -24:26:28.47 & $\rho$ Oph  &  84  & 120 & K4 & 2,7 \\
Flying Saucer      & 16:28:13.69, -24:31:39.48 & $\rho$ Oph  &  86  & 120 & M1 & 3,8 \\
ESO-H$\alpha$ 574  & 11:16:02.76, -76:24:53.24 & Cham I      &  80  & 192 & K8 & 4,7 \\
LkH$\alpha$ 263C          & 02:56:08.67, +20:03:40.60 & L1457       &  87  & 275 & M0 & 5 \\
OphE MM3           & 16:27:05.86, -24:37:08.36 & $\rho$ Oph  &  87  & 120 & K5 & 6
\enddata
\tablenotetext{}{(1) \citet{Stapelfeldt2003}, (2) \citet{Wolff2021} (3) \citet{Grosso2003} (4) \citet{Stapelfeldt2014} (5) \citet{Jayawardhana2002} (6) \citet{Brandner2000} (7) \citet{Villenave2020} (8) \citet{Guilloteau2016}}
\end{deluxetable*}

\subsection{Observations}
Imaging spectroscopy across the wavelength range from 1.66--27.9~$\mu$m was obtained using the NIRSpec and MIRI integral field units (IFUs) on JWST.  The NIRSpec/IFU observations, taken with the G235H/F170LP and G395H/F290LP grating/filter combinations, cover 1.66--5.27~$\mu$m with a spectral resolving power of $\sim$2700.  The MIRI/MRS observations, consisting of 4 channels each observed as 3 sub-bands (SHORT, MEDIUM, LONG), cover 4.90--27.9~$\mu$m with a spectral resolving power ranging from 1330 to 3710.  Most observations were taken as part of JWST Program \#5299 (JEDIce; PI: J.~Bergner).  The exception is MIRI/MRS observations of HV Tau C and Flying Saucer, which were available as part of JWST Program \#1282 (PI: T.~Henning; \citealt{Henning2024}).  Due to a failed guide-star acquisition, NIRSpec observations of OphE MM3 have not yet been obtained.

The NIRSpec observations were obtained using a cycling 12-point dither pattern using the NRSIRS2RAPID readout pattern, with 10 groups/integration (1926 seconds) for the G235H setting and 25 groups/integration (4552 seconds) for the G395H setting.  The G235H setting was mainly intended to constrain the red wing and continuum on the 3~$\mu$m water band, necessitating a lower signal-to-noise ratio than the weak ice features covered by the G395H setting. The NIRSpec observation of the Flying Saucer used a 2$\times$1-tile mosaic to fully cover its large angular extent, while all other disks were observed in a single tile.  

The MIRI observations obtained through Program \#5299 used a 4-point dither optimized for point sources, following target acquisition.  For each sub-band, 63 groups/integration and 2 integrations/dither were obtained with the FASTR1 readout pattern for a total exposure time of 1410 seconds.  HV Tau C and Flying Saucer were observed as part of Program \#1282 using a 4-point dither and no target acquisition.  The FASTR1 readout pattern was used with 27 groups/integration and 4 integrations/dither for a total exposure time of 1232 seconds per sub-band.  

\subsection{Data preparation}
\label{subsec:data_prep}
The MIRI observations were processed using the JWST Disk Infrared Spectral Chemistry Survey (JDISCS) pipeline described in \citet{Pontoppidan2024}.  The JWST calibration pipeline versions and calibration reference contexts used for each cube are listed in Table \ref{tab:calibration}.  One-dimensional MIRI spectra were extracted from the stage 2b cubes as described in \citet{Pontoppidan2024}.  

The NIRSpec cubes were reduced using the JWST calibration pipeline to stage 3. Outliers were detected and removed using a threshold of 90\% and kernel size of (9,9). One-dimensional NIRSpec spectra were extracted from the stage 3 cubes using a fixed elliptical aperture chosen for each disk to encompass the continuum emission at a $>$5$\sigma$ level (Table \ref{tab:calibration}).  Channels containing hot pixels that remained in the cubes after calibration were flagged by identifying outlier pixels, either from their absolute intensities or from the ratio of maximum to median intensities across all pixels within the channel.

All NIRSpec spectra presented here were background-subtracted based on the emission levels measured in the off-source regions of each cube. The JDISCS process uses pair-wise nod-subtraction to remove the extended background from the MIRI spectra.  Since the disks are extended, extractions of one-dimensional spectra are affected by various degrees of aperture-losses. To produce a continuous spectrum across the entire observed wavelength range, we adopted the G395H setting as our reference for absolute fluxes.  When needed, the fluxes of each neighboring spectral setting were scaled in turn, based on the median flux level measured within the overlapping wavelength range of adjacent settings.  The G235H fluxes were higher than the G395H fluxes by factors of 1--10\%.  The MIRI fluxes were lower than the G395H fluxes by 20--30\%, except for Oph 163131 (1.8$\times$ lower) and Flying Saucer (4.5$\times$ lower).

Due to a combination of narrow gas-phase emission and absorption lines along with broad ice-phase absorption features, special care was taken to estimate the continuum in each spectrum, using a method inspired by \citet{Pontoppidan2024}.  First, gas-phase lines were removed with the following procedure: over several iterations, (i) a median filter was applied to the spectrum, (ii) channels in which the flux deviated from the local median by more than a specified threshold were flagged, with the threshold for flagging decreasing each iteration, and (iii) flagged channels were filled in based on interpolation from the median-filtered spectrum.  After this, the spectrum was smoothed using Savitzky-Golay filtering.  For both the median filtering and the Savitzky-Golay filtering, kernel sizes in different wavelength ranges of the spectrum were adjusted to achieve a smoothly varying continuum without removing true ice absorption features.

\section{Program overview}
\label{sec:res}
In this section, we provide an overview of the data obtained through our program, including continuum images, full near-/mid-IR disk-integrated spectra, continuum-subtracted gas line spectra, and global optical depth spectra.  Our aim is to provide a holistic view of the information contained in the data cubes, highlight the most important features, illustrate the quality of the data, and identify avenues for deeper investigation in future work.

\begin{figure*}
    \centering
    \includegraphics[width=\linewidth]{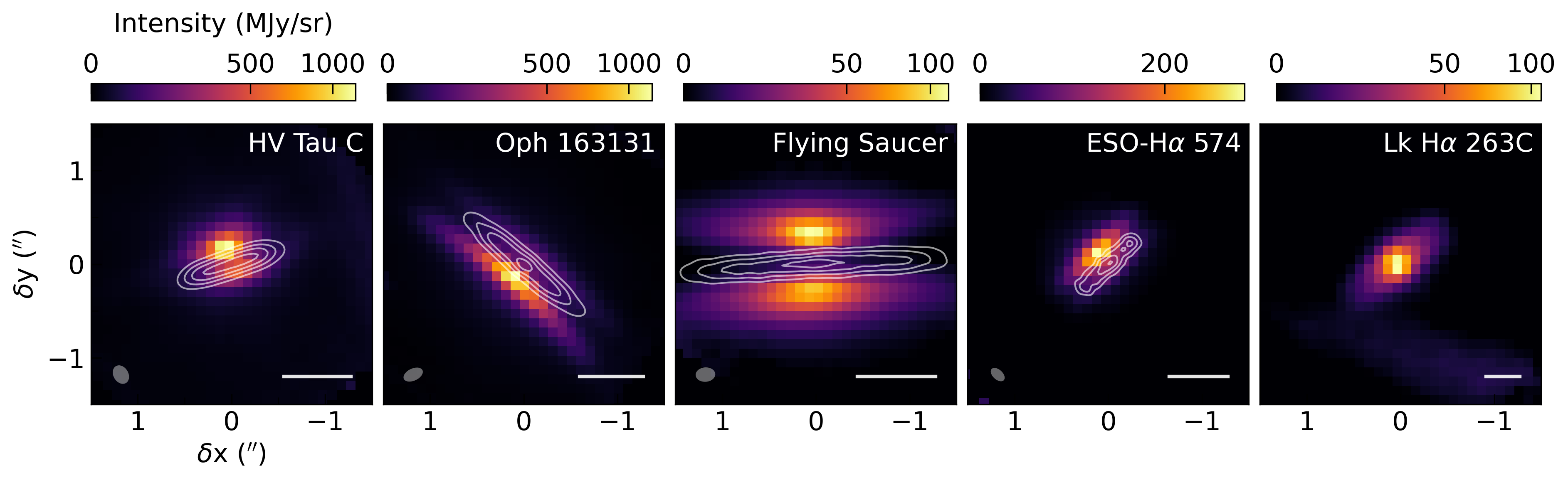}
    \caption{Intensity images show the near-IR continua observed with JWST, determined from the median intensity value of each spaxel across the G395H cube.  To visualize faint emission, color maps are shown with a power-law color stretch.  White contours show the millimeter continuum observed with ALMA (Band 6 or 7, see text for details), with contours drawn at [0.3, 0.5, 0.7, 0.9]$\times$ the peak intensity.  ALMA restoring beams are shown in the lower left of each panel, and white bars in the lower right represent linear scales of 100 au.}
    \label{fig:cont_images}
\end{figure*}

\subsection{Continuum images}
The near-IR continuum for each disk in our program is shown in Figure \ref{fig:cont_images}, along with the millimeter continuum observed with ALMA Band 6 or 7 when available.  The millimeter continuum data were retrieved from the ALMA Science Archive, and correspond to project codes 2023.1.00634.S (HV Tau C, 284 GHz), 2016.1.00771.S (Oph 163131, 224 GHz), 2023.1.00907.S (Flying Saucer, 347 GHz), and 2022.1.01302.S (ESO-H$\alpha$ 574, 225 GHz).  To mitigate small pointing discrepancies, the JWST and ALMA images were aligned by eye based on their continuum centers.

In the near-IR, most disks exhibit a clear two-lobed emission morphology separated by a dark lane at the midplane, due to the scattering of light along their upper and lower flared surfaces \citep{Padgett1999}.  The dark lane is not apparent for Lk H$\alpha$ 263C, possibly due to its much larger distance relative to the other disks, and/or a less inclined viewing inclination resulting in one dominant scattered light lobe.  The faint emission in the southwest of the Lk H$\alpha$ 263C image is associated with the nearby Lk H$\alpha$ AB binary.  The quality of the JWST observations will likely enable tighter constraints to be placed on the viewing inclinations of these disks compared to the literature values listed in Table \ref{tab:disk_sample}.  For instance, based on the symmetry of the upper and lower emission lobes, Flying Saucer appears closest to an edge-on viewing geometry, while all other sources exhibit one brighter lobe representing the surface angled towards the observer.  The lower lobe of ESO-H$\alpha$ 574 is particularly faint, consistent with a less-inclined geometry.  

In all cases, the millimeter continuum has a much smaller vertical extent than the near-IR continuum, as has been seen in previous comparisons of high-resolution millimeter and IR/optical observations of edge-on disks \citep{Villenave2020,Wolff2021,Villenave2024,Duchene2024,Tazaki2025,Dartois2025}.  This is explained by efficient vertical settling of large mm-sized grains to the midplane, while the small micron-sized dust remains well mixed in the disk upper layers.  The radial extents of the millimeter and near-IR continuum emission are more similar.  This does not necessarily imply that radial drift is inefficient: in Flying Saucer, the millimeter continuum is more radially compact than the IR continuum viewed \textit{in absorption} against the ambient field  \citep{Dartois2025b}, underscoring that the scattered IR photons do not necessarily trace the full radial extent of the small-dust disk.

\subsection{Disk-integrated spectra}
\begin{figure*}
    \centering
    \includegraphics[width=\linewidth]{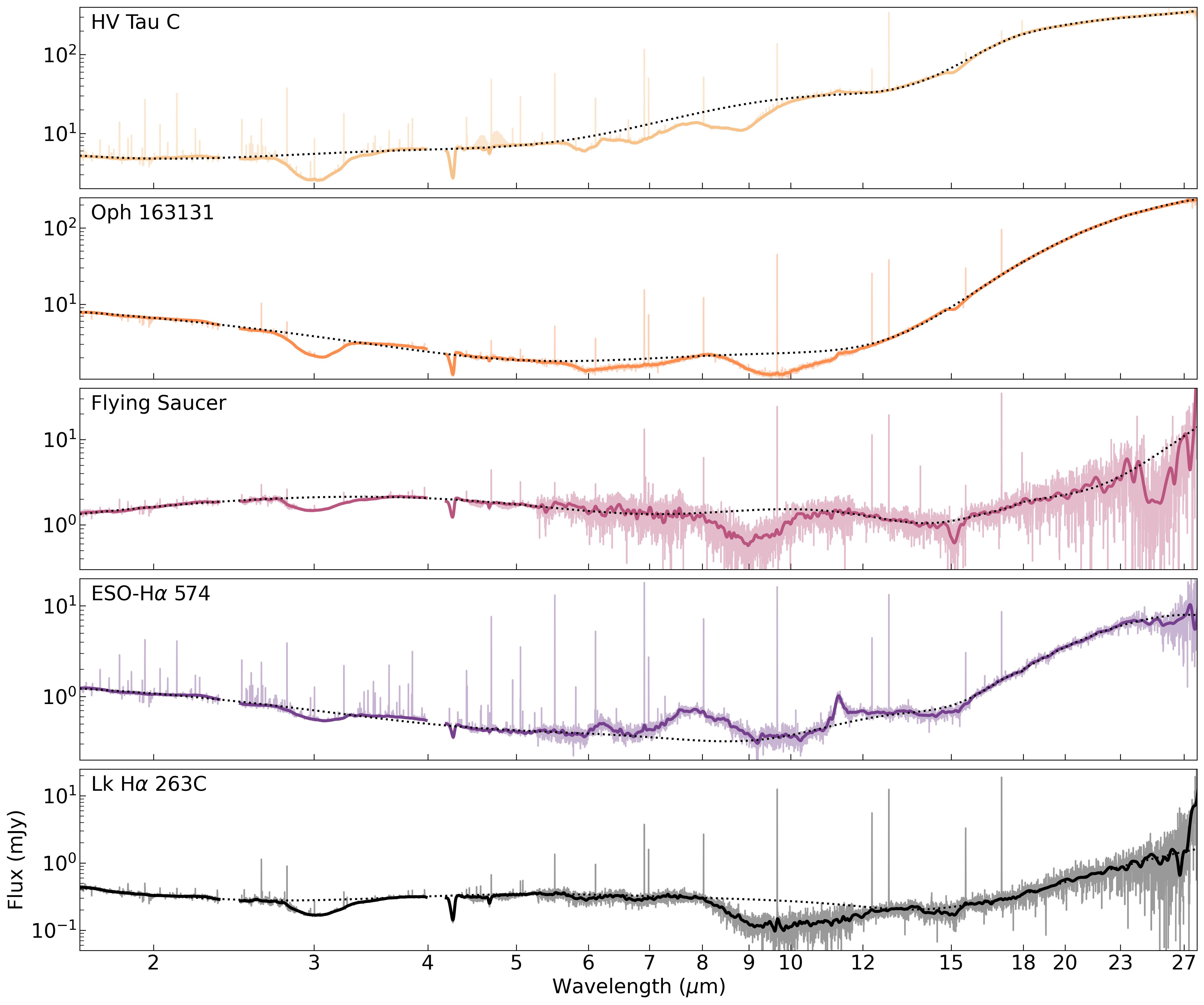}
    \caption{Disk-integrated NIRSpec/IFU + MIRI/MRS spectra for each disk.  Transparent lines show the original spectra, and the smoothed spectra are overlaid with opaque lines.  Dotted black lines show global continuum models produced by fitting a polynomial to feature-free regions of each spectrum (Table \ref{tab:continuum}).  Gaps around 2.4 and 4.1 $\mu$m are due to the G235H and G395H detector gaps.  }
    \label{fig:spec_nmir}
\end{figure*}

The disk-integrated spectra across the near- to mid-IR range are shown in Figure \ref{fig:spec_nmir}, extracted as described in Section \ref{subsec:data_prep}.  The signal-to-noise ratio is lower in the mid-IR than the near-IR, especially for sources with faint mid-IR emission.  The continuum emission of Flying Saucer, ESO-H$\alpha$ 574, and Lk H$\alpha$ 263C is relatively flat across the entire near-/mid-IR range, while HV Tau C and Oph 163131 show steeper flux increases at longer wavelengths, reflecting enhanced contribution from warm dust and possible differences in disk geometry. 

The near- to mid-IR spectra of all disks exhibit a wealth of emission and absorption features, which are discussed in more detail below.  Together, these features provide information on the disk ice inventories, polycyclic aromatic hydrocarbon (PAH) content, inner-disk gas composition, mass-loss tracers, and grain properties.  

\subsection{Gas emission \& absorption spectra}
\begin{figure*}
    \centering
    \includegraphics[width=\linewidth]{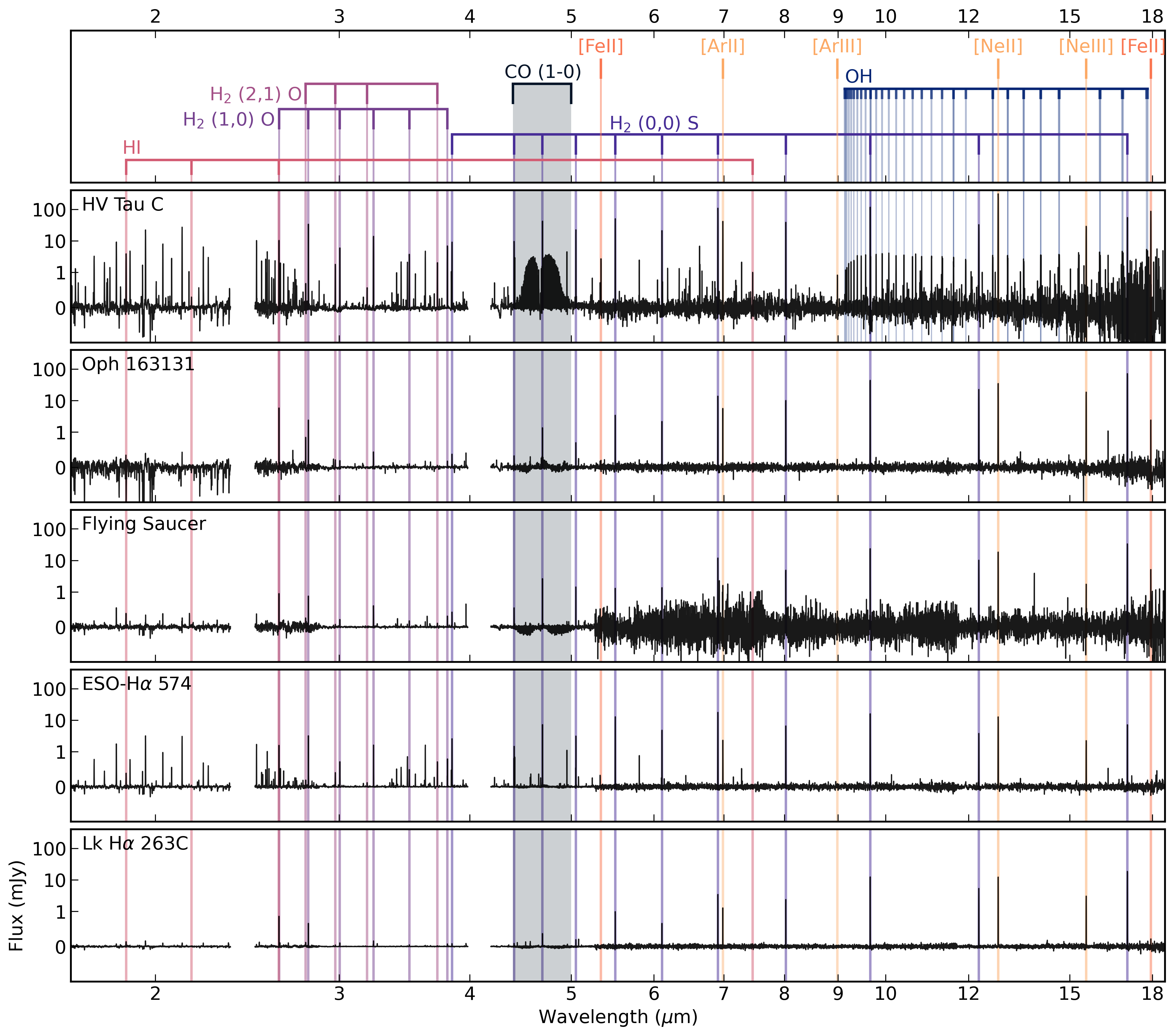}
    \caption{Continuum-subtracted disk-spectra highlighting the narrow gas-phase lines observed toward each disk. For reference, the top panel shows the positions of prominent atomic and molecular transitions detected towards our sample.}
    \label{fig:spec_gaslines}
\end{figure*}

\begin{figure*}
    \centering
    \includegraphics[width=\linewidth]{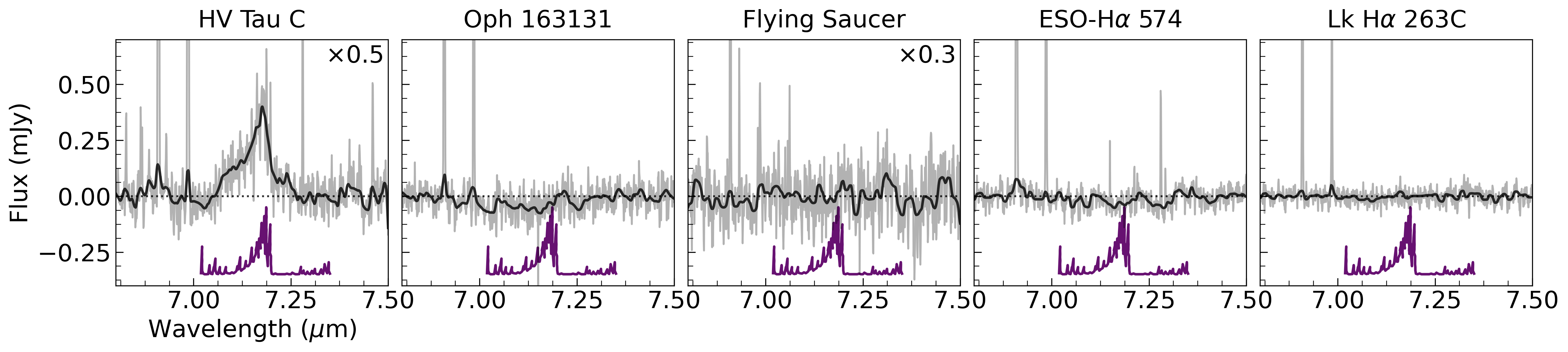}
    \caption{Zoom-in of the 7.2~$\mu$m region observed for our disks (grey lines show the original spectra, and black lines show the smoothed spectra), compared to a template of CH$_3^+$ emission (purple lines) taken from \citet{Changala2023}. For clarity, the observed spectra were scaled as indicated for several disks.}
    \label{fig:ch3p}
\end{figure*}

Numerous narrow emission and absorption lines from gas-phase species are evident in the spectra shown in Figure \ref{fig:spec_nmir}.  To compare these gas lines on the same scale across the sample, Figure \ref{fig:spec_gaslines} shows the same spectra following subtraction of the smoothed continuum (described in Section \ref{subsec:data_prep}) for each disk.  We focus on wavelengths $<$18.5~$\mu$m as there are no prominent lines detected beyond this and the spectra instead become dominated by noise.  Labels are included for prominent lines that are commonly detected towards our targets.  In general, HV Tau C shows the brightest and most line-rich gas-phase spectrum, followed by ESO-H$\alpha$ 574.  Oph 163131 and Flying Saucer exhibit fewer lines, but those detected are still relatively bright.  Lk H$\alpha$ 263C has a fairly weak and line-poor gas-phase spectrum.  Note that these spectra may exclude emission contributions from spatially extended jets beyond the extraction apertures.

Molecular hydrogen rovibrational transitions ($v$=0,1,2) from the S and O series are prominent throughout the entire near-/mid-IR range for all 5 disks in our sample.  Bright H$_2$ Q transitions are also seen around $\sim$2.5--2.6~$\mu$m for HV Tau C and ESO-H$\alpha$ 574, though for clarity these are not labeled in Figure \ref{fig:spec_gaslines}.  The Paschen $\alpha$ line at 1.875~$\mu$m is detected towards all disks, though in absorption towards Oph 163131.  Several additional H recombination lines are detected in HV Tau C, but they are not prominent in the other disks.  We also detect a number of forbidden atomic lines, of which [Ar~II] (6.98~$\mu$m), [Ar~III] (8.99~$\mu$m), [Ne~II] (12.81~$\mu$m), [Ne~III] (15.55~$\mu$m), and [Fe~II] (5.34, 17.93~$\mu$m) are among the strongest.   While beyond the scope of the present work, the detection of the noble gas lines towards all 5 disks will provide constraints on the radiation field at the disk surfaces \citep{Hollenbach2009,Pascucci2020}, while the morphologies of the Ne and Fe transitions can provide insight into the nature of the disk winds and mass loss mechanisms \citep{Arulanantham2024,Bajaj2024,Pascucci2025,Kurtovic2026}.  All disks also exhibit a series of narrow absorption lines from $\sim$1.5--2~$\mu$m, which may be features from the stellar photosphere seen in the scattered light.  

The gas-phase CO $v$=1--0 ro-vibrational ladder is also seen towards all disks, consisting of transitions with upper-state energies spanning $\sim$3000--8000 K.  In previously observed edge-on disks, CO ro-vibrational emission is reported to be either scattered out from the hot inner disk, or tracing outflowing wind material \citep{Sturm2023c, Arulanantham2024,Pascucci2025}.  In our sample, the CO lines are most intense towards HV Tau C, which is also the only source where they are primarily seen in emission.  The lines are mainly in absorption towards Flying Saucer and Lk H$\alpha$ 263C, and in a combination of emission and absorption towards Oph 163131 and ESO-H$\alpha$ 574.  

Several gas-phase species are detected towards HV Tau C that are not seen towards the other disks.  Notably, we detect numerous OH emission lines towards HV Tau C, including the `prompt' emission lines at $\sim$9--12$\mu$m.  These correspond to emission from excited levels populated via photodissociation of H$_2$O \citep{Tabone2021,Zannese2024,Neufeld2024,Watson2025} and indicate that a UV-dominated and water-rich environment is being traced.  We do detect weak lines at several positions around 14 $\mu$m where H$_2$O rotational transitions are expected, but spectral modeling is needed to confirm whether the overall emission pattern is consistent with H$_2$O.  HV Tau C also shows signatures of the CH$_3^+$ molecular ion around 7.2 $\mu$m (Figure \ref{fig:ch3p}), which has now been detected towards several disks in both irradiated and quiescent environments \citep{Berne2023,Henning2024,Schroetter2025,Romero2025,Volz2026}.  This further supports that HV Tau C is host to an especially active gas-phase photochemistry, though it remains unclear exactly what environment it originates from (e.g.~disk surface, wind, etc.)

It is clear from this overview that the gas-phase spectra contain a tremendous amount of information about the chemical, physical, and dynamical environment of these disks.  In future work, a more detailed analysis of the gas lines -- including their emission morphologies, line kinematics, and excitation conditions -- will help to reveal the quantity and physical origins of the atomic and molecular gases probed by our observations.

\subsection{Global optical depth spectra}

\begin{figure*}
    \centering
    \includegraphics[width=0.95\linewidth]{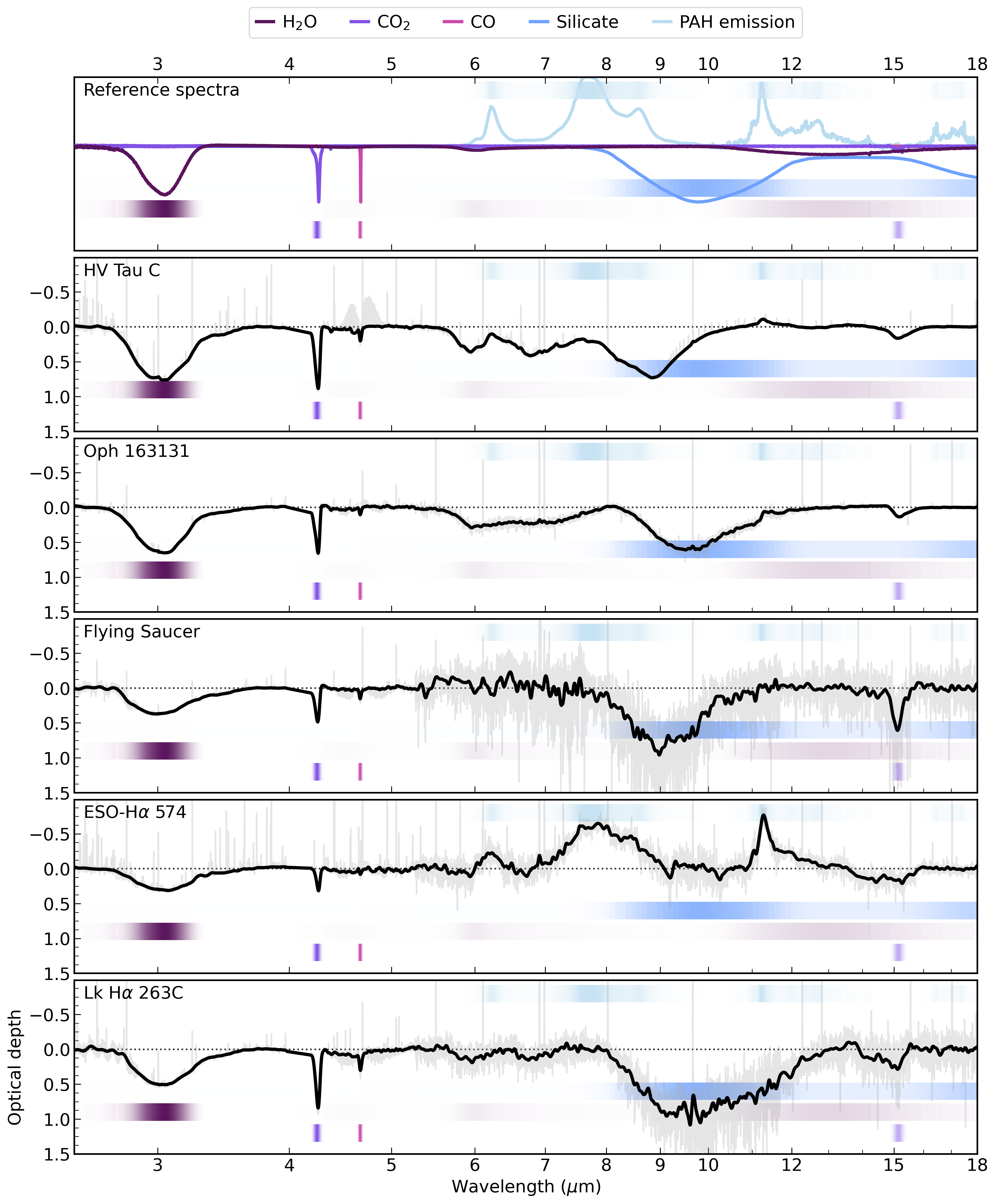}
    \caption{Global optical depth spectra for each disk, determined using the continuum models shown in Figure \ref{fig:spec_nmir}.  Grey lines show the original spectra, and black lines show the smoothed spectra.  For reference, the top panel shows the absorption profiles of major ice and silicate species and the emission profile of a PAH template.  In all panels, the relative intensity as a function of wavelength for these reference species is represented via transparency of the colored horizontal bars.}
    \label{fig:spec_refbands}
\end{figure*}

To highlight the broad features that contribute to the overall spectral shape, Figure \ref{fig:spec_refbands} shows a global optical depth spectrum for each disk, made by fitting a polynomial continuum (shown in Figure \ref{fig:spec_nmir}) to featureless regions across the entire near- to mid-IR (Table \ref{tab:continuum}).  For comparison, we also show spectral templates for the PAH emission and ice absorption features that are readily identified in the global optical depth spectrum.  The low signal-to-noise ratio at mid-IR wavelengths for Flying Saucer and, to a lesser extent, Lk H$\alpha$ 263C, makes it somewhat more challenging to interpret the depth and shape of mid-IR absorption features in these disks.  Additionally, note that in some cases the smoothed optical depth spectra show small artifacts due to imperfect mitigation of bright spectral lines (e.g.~near the 7~$\mu$m H$_2$ and [ArII] lines).

The PAH template corresponds to the spectrum of the HD 97300 disk observed with \textit{Spitzer} and shown in \cite{Dartois2025}.  The strong PAH features centered at 6.2, 7.7, and 11.3~$\mu$m dominate the JWST spectrum of ESO-H$\alpha$ 574.  ESO-H$\alpha$ 574 also exhibits fainter emission from the aromatic and aliphatic carbon features at 3.3 and 3.4~$\mu$m.  Of the remaining disks, the 11.3~$\mu$m feature can be clearly identified towards HV Tau C and Oph 163131, but is not seen towards Flying Saucer or Lk H$\alpha$ 263C.  Note that a PAH background is seen towards Flying Saucer \citep{Dartois2025b}, but there is no excess PAH emission at the disk position.  Additional PAH features may contribute at other wavelengths, but are more difficult to identify due to overlap with ice and silicate bands and increasing extinction at shorter wavelengths \citep[e.g.][]{Sturm2024}.  

Prior to JWST, PAH emission had not been detected towards low-mass (M/K-type) T Tauri stars \citep{Geers2006}.  With the three detections in our sample, along with Tau 042021 and HH 48 NE \citep{Arulanantham2024,Sturm2024}, there are now five detections of PAH emission localized to the disks around low-mass T Tauri stars.  This supports that PAH emission is not in fact absent from these low-mass systems, though work remains to understand why it is only detected for edge-on viewing geometries.

Ice features are detected towards all disks in our sample.  Figure \ref{fig:spec_refbands} highlights the absorption bands from the abundant species H$_2$O, CO$_2$, and CO.  The transmission spectra shown for reference all correspond to 15 K pure ices and were taken from \citet{Oberg2007} (H$_2$O) and \citet{vanBroekhuizen2006} (CO$_2$, CO) via the Leiden Ice Database for Astrochemistry (LIDA; \citealt{Rocha2022}). The H$_2$O OH stretch at 3~$\mu$m is prominent in all 5 disks, while the broader and weaker libration mode around 13~$\mu$m is not readily distinguishable from the continuum.  Several disks exhibit apparent absorption in the 6~$\mu$m region around the H$_2$O bending mode, but other species may also contribute to absorption at similar wavelengths (discussed further in Section \ref{subsec:ice_6um}).  The CO$_2$ stretch at 4.27~$\mu$m is the deepest ice band observed towards most disks.  The neighboring $^{13}$CO$_2$ stretch at 4.38~$\mu$m is clearly apparent in HV Tau C, Oph 163131, and Lk H$\alpha$ 263C, and tentatively seen in Flying Saucer and ESO-H$\alpha$ 574 (Figure \ref{fig:13co2}).  The 15~$\mu$m CO$_2$ bend is weaker, but still detected towards all 5 disks.  Lastly, the CO stretch at 4.67~$\mu$m is seen towards all 5 disks.

Silicate absorption also contributes to the shape of the global optical depth spectrum. While the exact position, shape, and depth will depend sensitively on the grain properties \citep[e.g.][]{Henning2010,Sturm2023b,Arabhavi2022}, we show for illustration the absorption profile of a silicate population spanning 0.005--5~$\mu$m in size, generated with \texttt{optool} \citep{Dominik2021} with optical constants from \citet{Dorschner1995} for a pyroxene stoichiometry. The deep absorption band around 9--11~$\mu$m is clearly seen towards all disks except ESO-H$\alpha$ 574, whose mid-IR spectrum is dominated by PAH emission.  The silicate feature varies in position, shape, and optical depth across the sample, hinting at diversity in the grain populations of these disks, though radiative transfer effects could also be contributing.  

The empirical identification of a global continuum is challenging, particularly in the mid-IR, where there is overlap between broad ice and silicate absorption features and PAH emission features.  However, the choice of continuum model does not affect our conclusions about what features are present in the disk spectra.  

\section{Ice survey results}
\label{sec:ice}
Our program provides the first opportunity to carry out a comparative analysis of the ice features observed across a sample of edge-on disks.  We first discuss the major ice species H$_2$O, CO$_2$, and CO, which are detected unambiguously towards all disks in our survey, and then consider evidence for the presence of lower-abundance ice species.  Since radiative transfer effects may perturb the ice spectral features observed in edge-on disks \citep[including the absolute optical depths of ice bands, the shape of the band profiles, and the relative depths of bands from the same molecule that occur at different wavelengths; ][]{Pontoppidan2005,Dartois2022,Arabhavi2022,Sturm2023b,Sturm2023c,Bergner2024,Martinien2025}, we focus here on what can be learned with an empirical approach.  Radiative transfer modeling to constrain the ice abundances and ice mixture components will be presented in future work.

\subsection{H$_2$O, CO$_2$, and CO ice bands}
\label{subsec:ice_major}

\subsubsection{Absolute optical depths}

\begin{figure*}
    \centering
    \includegraphics[width=\linewidth]{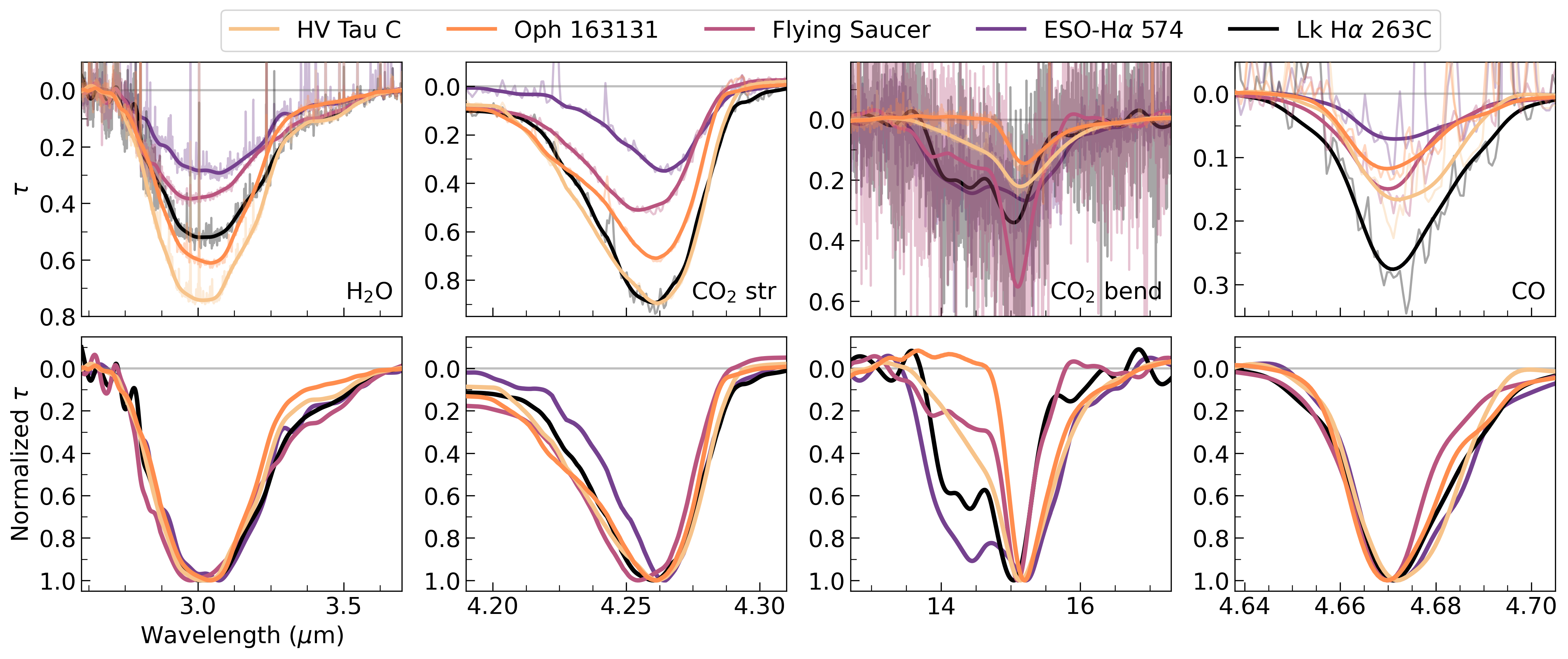}
    \caption{Top: local optical depth spectra for the main H$_2$O, CO$_2$, and CO ice bands observed towards the disk sample.  Original spectra are shown with transparent lines, and smoothed spectra are overlaid with opaque lines.  Bottom: as above, but showing normalized optical depth profiles for the smoothed spectra only.}
    \label{fig:ice_major}
\end{figure*}

The optical depth spectra of the main H$_2$O, CO$_2$, and CO ice bands are shown in Figure \ref{fig:ice_major} (top).  These are calculated from  $-\mathrm{ln}(F/F_\mathrm{cont})$, where $F$ is the observed flux and $F_\mathrm{cont}$ is the local continuum model.  Note that this observable is not a true line-of-sight optical depth due to radiative transfer effects, and therefore cannot be directly converted to a line-of-sight ice column density \citep[e.g.][]{Sturm2023c}.  The baseline regions and continuum flux models used to produce these local optical depth spectra are shown in Figure \ref{fig:local_cont}. 

The peak optical depths range from $\sim$0.3--0.8 for H$_2$O, 0.3--0.9 for the CO$_2$ stretch, and 0.05--0.3 for CO.  For most disks, the CO$_2$ bend has a peak optical depth around 0.1--0.3.  Flying Saucer is the exception, with an apparent optical depth around 0.5, though the signal-to-noise ratio is low.  To a lesser extent, ESO-H$\alpha$ 574 and Lk H$\alpha$ 263C also suffer from low signal-to-noise around the CO$_2$ bending mode, and deeper observations would help to confirm these feature shapes and depths.  

We now consider the relative ordering of the ice band depths, focusing on the near-IR bands for which signal-to-noise is not a concern.  ESO-H$\alpha$ 574 has the smallest optical depth for all 3 bands, and the optical depths are also generally small towards Flying Saucer.  HV Tau C exhibits the deepest H$_2$O and CO$_2$ stretching bands, and a relatively deep CO band.  Oph 163131 has relatively deep H$_2$O and CO$_2$ bands, but the second-shallowest CO band.  Lk H$\alpha$ 263C has a modest H$_2$O band depth, one of the deepest CO$_2$ bands, and a CO band nearly twice as deep as any other disk.  

Together, this shows that our sample contains a fairly wide range of absolute optical depths for a given ice band (varying by factors of $\sim$3--6), as well as relative optical depths of different ice bands.  This variation is suggestive of a range in the ice richness and composition of the different disks in the sample.  We can speculate about the significance of these optical depth trends based on the modeling presented in \citet{Sturm2023b}.  In that work, the apparent ice optical depths were found to be only weakly sensitive to variations in  disk structure parameters (characteristic radius, flaring index), as well as to variations in the inclination for values $>$70$^\circ$.  Ice optical depths were found to be modestly sensitive to the dust grain properties, and to show the strongest dependency on the ice abundances and disk mass.  Together, this indicates that the observed variations in the optical depths of ice features in our sample may be due to true differences in the ice abundances, but could also reflect differences in the disk masses or, to a lesser extent, dust grain properties.  Radiative transfer modeling will be required to draw definitive conclusions about the absolute ice abundances \citep{Pontoppidan2005,Sturm2023c, Sturm2023b}.  On the other hand, the absorption features from different molecules should respond in a similar way to changes in the disk mass.  The observed variations in the relative optical depths of different ice bands is therefore likely to trace real variations in the ice-phase molecular mixing ratios across our sample.

\subsubsection{Ice band profiles}
Figure \ref{fig:ice_major} (bottom) shows the normalized ice optical depth profiles, i.e.~$\tau$/$\tau_\mathrm{max}$, to enable a comparison of the absorption band profiles.  For clarity, we show only the smoothed spectra determined as for the global continuum (Section \ref{subsec:data_prep}).  The shapes of the H$_2$O, CO$_2$ stretch, and CO bands are approximately similar across the sample, with some variation in the wings.  In all disks, H$_2$O exhibits a strong absorption peaking around 3.0~$\mu$m, though the peak appears slightly blue-shifted in Flying Saucer and red-shifted in Oph 163131 relative to the other disks.  All disks also exhibit a shoulder of varying relative intensity extending to $\sim$3.6~$\mu$m, which is most prominent for Flying Saucer and least for Oph 163131 and HV Tau C.  This red wing has been seen along protostellar lines of sight for decades, and is usually attributed to a combination of scattering by large icy grains, NH$_3$ hydrates, and possibly C-H stretches arising from hydrocarbons and methanol \citep[e.g.][]{Leger1983,Smith1989,Dartois2001,Dartois2002,Gibb2004}.  Flying Saucer also shows a prominent shoulder at 2.83 $\mu$m, which may be due to scattering and ammonia hydrates as well.  Note that the wiggles in the baseline below $\sim$2.75~$\mu$m in several disks reflect the lower signal-to-noise of the G235H setting.  

The CO$_2$ stretching band centers overlap exactly for HV Tau C, Oph 163131, and Lk H$\alpha$ 263C, while the band peak for Flying Saucer is somewhat blue-shifted and for ESO-H$\alpha$ 574 is somewhat red-shifted.  For all disks, the red side of the band is nearly overlapping, and the blue side exhibits a broad shoulder that varies somewhat in shape and amplitude. ESO-H$\alpha$ stands out for having a particularly narrow shape; while it is still asymmetric, the blue shoulder is much less prominent than in the other disks.  The absolute optical depth of this band is also very low, so the narrow profile could reflect a lower degree of saturation.  While the CO$_2$ stretching band profile has been used for decades to determine the thermal history and mixing status of CO$_2$ ice along interstellar lines of sight \citep[e.g.][]{Gerakines1999}, in disks it is subject to saturation and radiative transfer effects \citep{Sturm2023b,Bergner2024}, and modeling will be necessary to interpret the observed variations in band shape across the sample.

The CO$_2$ bending region appears to show two distinct peaks, one centered at the CO$_2$ bending position of 15.2~$\mu$m, and one around 14.2~$\mu$m.  Both peaks are comparable in strength towards ESO-H$\alpha$ 574 and LkH$\alpha$ 263C; towards HV Tau C and Flying Saucer the 14.2~$\mu$m feature is weaker, appearing as a shoulder; and it is not present at all towards Oph 163131.  The possible identity of the 14.2~$\mu$m band is discussed further in Section \ref{subsec:ice_trace}.  Focusing on the CO$_2$ bend, all disks exhibit a very similar peak position, though Lk H$\alpha$ 263C is slightly blue-shifted relative to the others.  HV Tau C, Oph 163131, and ESO-H$\alpha$ 574 all have broad red shoulders with a nearly identical profile, while Flying Saucer and Lk H$\alpha$ 263C are much narrower.  In protostars, such differences are attributed to an apolar CO$_2$ component producing the narrow feature, and a polar H$_2$O:CO$_2$ mixture contributing a broader, red-shifted component \citep{Ehrenfreund1997, Pontoppidan2008,Brunken2025}, though it is not clear from an empirical analysis alone whether this is true for our disks.  Interestingly, while the CO$_2$ stretching mode for ESO-H$\alpha$ 574 is a clear outlier among the sample, its CO$_2$ bending mode appears very typical.  This suggests that these features may be tracing different regions within the disk, as was found from radiative transfer modeling of HH 48 NE \citep{Sturm2024}.  

\begin{figure}
    \centering
    \includegraphics[width=\linewidth]{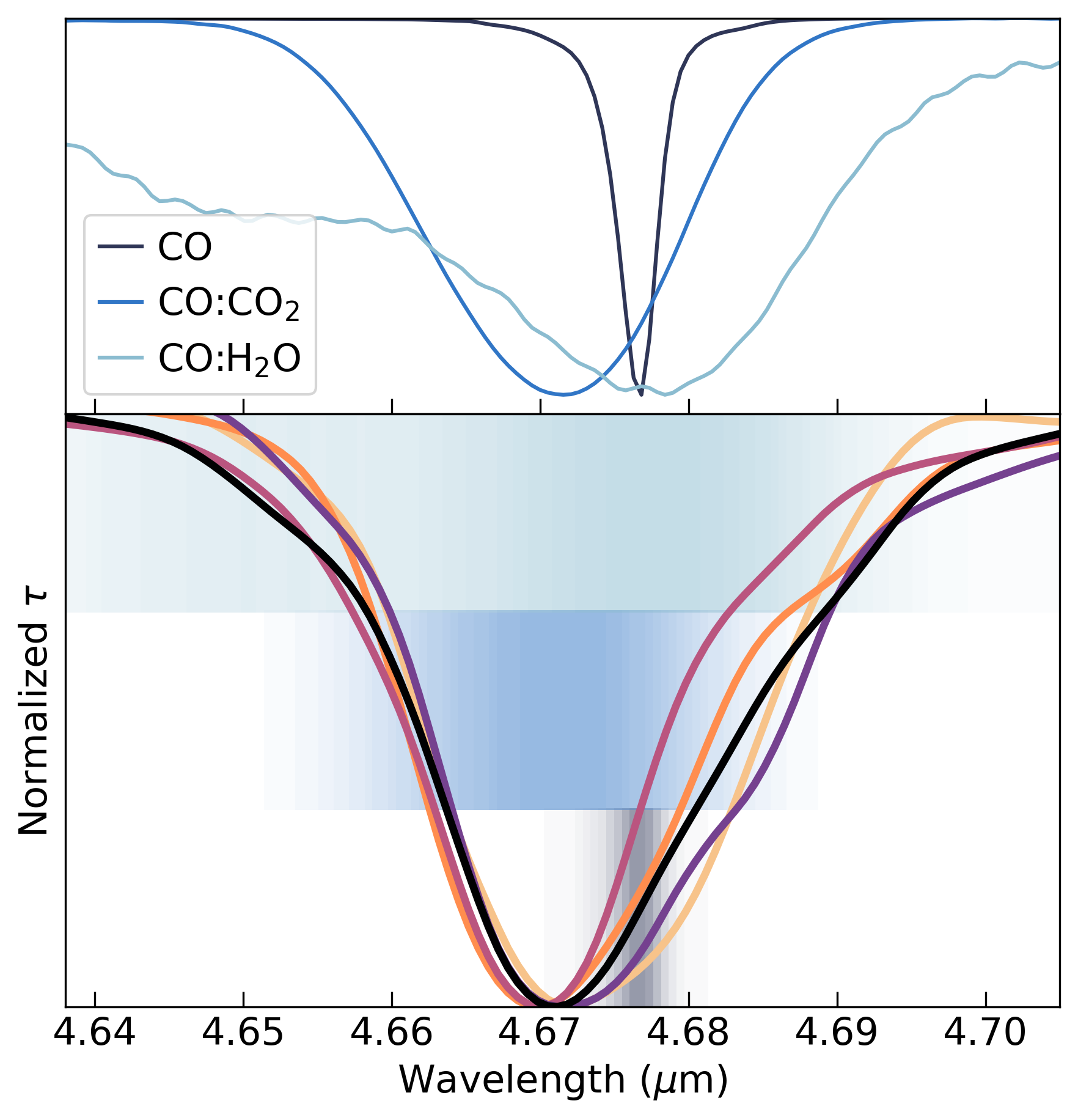}
    \caption{Top: lab-measured absorption spectra for pure CO, a 1:1 CO:CO$_2$ mixture, and a 1:10 CO:H$_2$O mixture.  All spectra were measured at 10 K and are taken from \citet{Bergner2024}.  Bottom: absorption profiles for the different lab mixtures are represented via transparency of the colored horizontal bars, and compared to the normalized optical depth profiles for all disks in our survey.}
    \label{fig:zoom_CO_ice}
\end{figure}
The CO band profiles are well aligned on the blue side and share similar peak position across the sample.  There is some variation in the shape of the red shoulder, which is narrowest for Flying Saucer and broadest for HV Tau C and ESO-H$\alpha$ 574.  Compared to CO$_2$ and H$_2$O, the CO band profile is much less sensitive to radiative transfer effects in edge-on disks, with models showing only slight differences  compared to thin film band profiles \citep{Bergner2024}.  Its observed shape therefore more reliably encodes information about the CO mixing environment \citep[e.g.][]{Tielens1991}.  Figure \ref{fig:zoom_CO_ice} compares the observed CO band profiles across our sample to the lab-measured spectral profiles of pure CO ice, CO mixed with CO$_2$, and CO mixed with H$_2$O, with all ices formed at 10~K and spectra measured at 10~K \citep{Bergner2024}.  The apolar CO:CO$_2$ mixture clearly provides the best match to the widths and peak positions of the observed CO band profiles.  By contrast, the CO:H$_2$O mixture is much broader and the pure CO ice is much narrower than the observed CO bands, and both exhibit a somewhat red-shifted peak position relative to the observed bands.  Therefore, our observations appear to predominantly trace CO ice in an apolar mixture with CO$_2$.  Similarly, previous modeling revealed that the majority ($\sim$70\%) of CO ice in the HH 48 NE disk is in a CO:CO$_2$ mixture \citep{Bergner2024}.  CO ice dominated by an apolar mixture is thus a common feature of the edge-on disks observed by JWST so far.  

The presence of CO:CO$_2$ ice mixtures may have important implications for how CO is partitioned between the gas and solid phases across the disk.  CO can be trapped within CO:CO$_2$ mixtures \citep{Simon2019}, and efficient transport of entrapped CO interior to its snowline would significantly alter the C/O ratios of planetary building blocks  \citep{Bergner2024, Ligterink2024, Williams2025}.  The ubiquity of mixed CO ice in all disks observed to date indicates that there is widespread potential for its entrapment.

\subsubsection{Optical depth maps}
\label{subsec:ice_maps}

\begin{figure*}
    \centering
    \includegraphics[width=\linewidth]{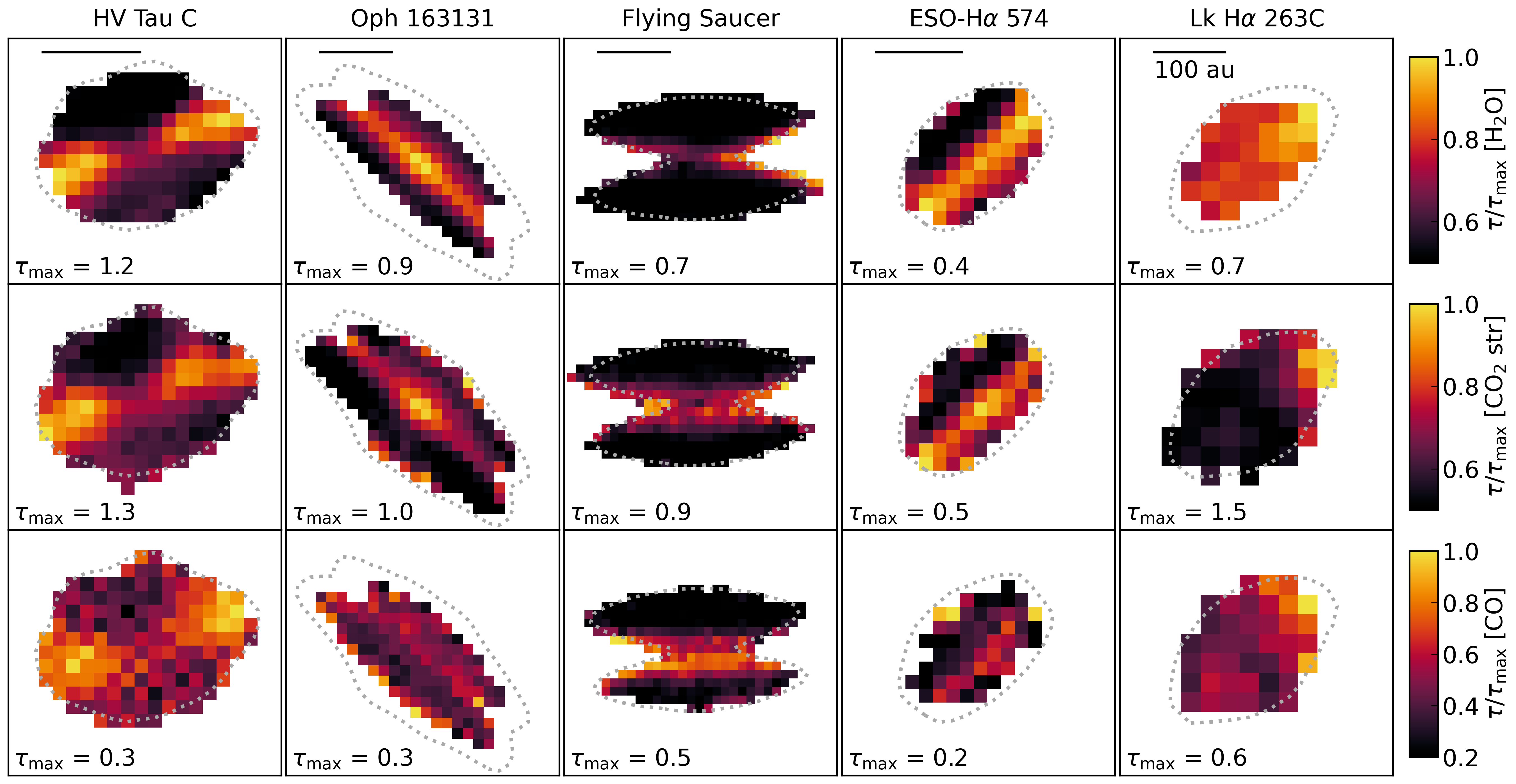}
    \caption{Optical depth maps of the near-IR H$_2$O, CO$_2$, and CO ice features in the disk sample, calculated for pixels with a $>$3$\sigma$ detection across the ice band.  For clarity, the color map within each panel is individually normalized (i.e., divided by the maximum per-pixel optical depth for a given ice band in a given disk).  The maximum optical depths are indicated in the bottom left of each panel. Dotted grey lines represent the 3$\sigma$ contour of the near-IR continuum maps shown in Figure \ref{fig:cont_images}. Horizontal bars represent a linear scale of 100 au.}
    \label{fig:ice_maps}
\end{figure*}
With the NIRSpec IFU, it is possible to spatially resolve the ice features observed towards edge-on disks.  Figure \ref{fig:ice_maps} shows the normalized peak optical depth maps for the near-IR H$_2$O, CO$_2$, and CO bands.  For a given ice band, optical depths were measured for all pixels where the median flux is at least 3$\times$ the off-source noise level within the wavelength range of the band, for which we adopt 2.68--3.60~$\mu$m for H$_2$O, 4.19--4.33~$\mu$m for CO$_2$, and 4.65--4.71~$\mu$m for CO.  While ice may be present in other regions, the signal-to-noise is too low to determine the depth of any absorption features.  For each pixel with sufficient SNR, local optical depth spectra were calculated similar to those shown in Figure \ref{fig:ice_major}, and the peak optical depth was determined across the band.  The peak optical depth in each pixel was divided by the highest value across the image in order to simultaneously visualize patterns across all disks and all bands.

For all disks, there is a measurable optical depth for each ice band across virtually the entire continuum region.  In HV Tau C, the optical depth increases vertically towards the midplane and radially away from the disk center.  Oph 163131, Flying Saucer, and ESO-H$\alpha$ 574 show a strip of higher optical depth along the entire midplane.  In some cases (H$_2$O in Oph 163131 and CO$_2$ in Oph 163131 and ESO-H$\alpha$ 574), there is a peak in optical depth at the disk center.  The optical depth maps of Lk H$\alpha$ 263C do not show any clear spatial gradients except for a possible increase in optical depth towards the northwest side of the disk.  This lack of midplane spatial features is possibly due to the fact that this disk is further away than others in our sample (Table \ref{tab:disk_sample}), or could indicate a contribution of foreground cloud material.  

The optical depth maps for the disks in our program show important similarities with the previously observed edge-on disks HH 48 NE and Tau 042021 \citep{Sturm2023c,Sturm2024,Dartois2025}.  As in those cases, we detect the major ice features out to large vertical separations from the disk midplane, in nearly all regions where there is a sufficient continuum signal to detect the ice absorption.  Radiative transfer modeling of HH 48 NE and Tau 042021 demonstrated that the vertical distribution of ice absorption features is not a scattering artifact, but requires ices to be present at high elevations \citep{Sturm2024,Dartois2025}.  This indicates that either the atmosphere is surprisingly well-shielded from UV photons, or that robust vertical mixing continuously supplies the upper layers with icy grains \citep[e.g.][]{Woitke2022}.  With now a total of 7 disks to compare, it is clear that elevated icy grains are the norm for protoplanetary disks.

While vertically extended ice features appear universal, we are beginning to see diversity in the radial optical depth patterns from disk to disk.  The optical depth maps of HV Tau C, which reach a midplane radial minimum at the disk center, are quite similar to those seen in HH 48 NE \citep{Sturm2023c}.  However, in other cases the opposite pattern is seen, in which the optical depths reach their maxima in the midplane center, and in some cases the optical depths are essentially flat across the entire midplane.  Whether these different midplane radial profiles are caused by radiative transfer effects or differences in the ice spatial distributions currently remains unknown, and source-specific radiative transfer modeling is needed to better understand their origins.

In most cases, the H$_2$O, CO$_2$, and CO ice bands show a similar spatial pattern within a given disk, indicating either that the ices are co-distributed in the disk or that the features are being uniformly scattered out from a smaller region.  Still, in a few cases, we can see differences in the optical depth maps for the different molecules.  Flying Saucer presents the best opportunity for this as its midplane region is well-resolved and the ice features in the near-IR range are generally detected at high signal-to-noise.  The H$_2$O optical depth reaches its maximum within a conical region tracing the edges of the scattering surface.  In contrast, the CO$_2$ and CO optical depths are flat across the entire midplane.  While the CO$_2$ optical depth gradually decreases with elevation from the midplane, CO exhibits a sharp transition from a high optical depth ($\sim$0.4) to a low optical depth ($\sim$0.1) at a projected elevation of 0.3$''$.  The fact that the CO optical depth does not drop to zero beyond this transition confirms that it is not an artifact due to low signal-to-noise.  One intriguing possibility is that the sharp CO transition could be a signature of the CO snow surface, which may be viewable for this disk alone thanks to its large size and almost perfect edge-on viewing geometry.  In any case, it is clear that in some disks there are spatial variations in the different ice features, which must reflect either underlying differences in the spatial distribution of the ices or distinct radiative transfer outcomes for the different absorption features.  

\subsection{Trace ice species}
\label{subsec:ice_trace}

\begin{figure*}
    \centering
    \includegraphics[width=\linewidth]{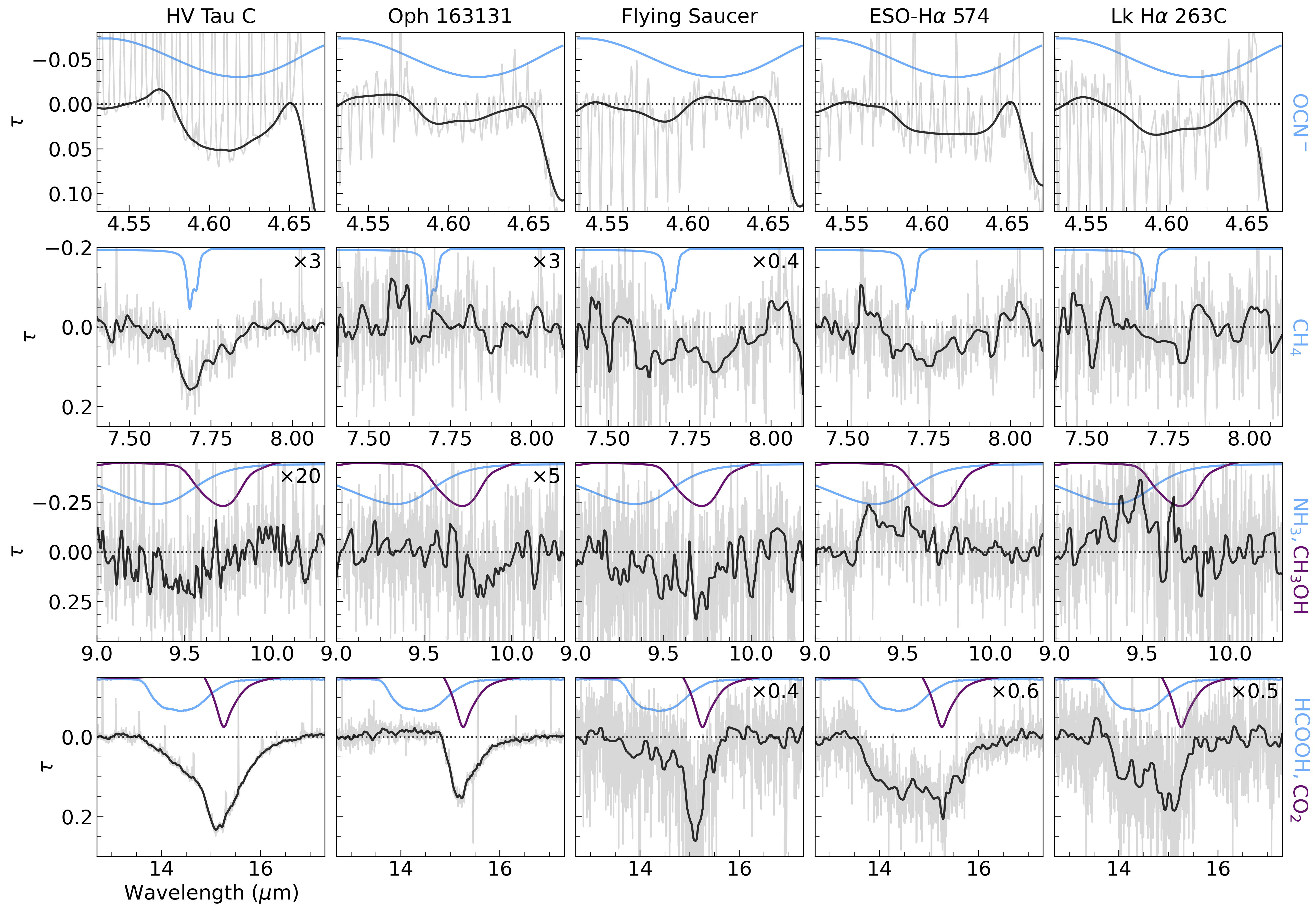}
    \caption{Local optical depth spectra in regions containing features from low-abundance ice species.  Colored lines show reference laboratory spectra for trace ice candidates.  Grey lines show the original spectra, and black lines show the smoothed spectra.  Laboratory spectra are taken from \citet{Novozamsky2001} (OCN$^-$), \citet{Rachid2020} (CH$_4$), \citet{Taban2003} (NH$_3$), \citet{Terwisscha2018} (CH$_3$OH), \citet{Hudson2024} (HCOOH), and \citet{Ehrenfreund1999} (CO$_2$) using LIDA \citep{Rocha2022}.}
    \label{fig:ice_trace}
\end{figure*}

In addition to the major ice features from H$_2$O, CO$_2$, and CO, the near-/mid-IR range contains absorption features from numerous additional ice components that may be present at lower abundances.  Figure \ref{fig:ice_trace} shows a selection of spectral regions where such features are expected.  The continuum models used to produce these optical depth spectra are shown in Figure \ref{fig:local_cont}.  For comparison, we also show lab-measured reference spectra, though note that the exact shape and position of these bands will depend on the ice environment like mixing conditions and temperature.  Our intent is to discuss likely carriers, while detailed spectral analysis using a wider range of templates will be the subject of future work.  Additionally, radiative transfer modeling will be needed to quantify the abundances or upper limits of these ice species, and to compare the depths of bands from the same molecule observed at different wavelengths \citep{Pontoppidan2005,Sturm2023b}.  Table \ref{tab:ice_dets} summarizes the detection status of each ice feature across the disk sample.

\subsubsection{4.6 $\mu$m}
OCN$^-$ exhibits a prominent absorption feature around 4.6 $\mu$m, which has been previously reported in the HH 48 NE disk \citep{Sturm2023c} and is commonly detected in protostellar ices \citep{vanBroekhuizen2005,Nazari2024}.  Although this spectral region overlaps with the CO ro-vibrational R branch,  for all disks except Flying Saucer we still discern a broad dip in the continuum compatible with OCN$^-$.  The feature is deepest for HV Tau C, with a peak optical depth of $\sim$0.05, followed by ESO-H$\alpha$ and Lk H$\alpha$ 263C, and shallowest towards Oph 163131.  As this feature lacks other compelling carriers and has been previously assigned to OCN$^-$ in a disk, we consider OCN$^-$ to be responsible for the 4.6 $\mu$m absorption seen in 4 of our 5 disks.  This makes OCN$^-$ the most common of the trace ice species that we could identify in our survey.

\subsubsection{7.7 $\mu$m}
A narrow absorption feature around 7.7 $\mu$m is detected towards HV Tau C, with no other disks showing clear signs of absorption at this position.  A likely candidate for the feature seen in HV Tau C is CH$_4$, which is shown for reference, though SO$_2$, OCN$^-$, and N$_2$O are also suggested to contribute to this region in protostellar ices \citep{Boogert1997, Oberg2008, Rocha2025, Gross2025,Karteyeva2026}.  As OCN$^-$ and SO$_2$ exhibit peaks somewhat red-shifted from the observed absorption (around $\sim$7.5--7.6~$\mu$m), we consider them unlikely to be the main carriers.  A weak absorption ($\tau \; <$ 0.03) in this region was also detected in HH 48 NE and attributed to an ISM-like CH$_4$/H$_2$O abundance of $<$12\% \citep{Sturm2024}.

\begin{deluxetable}{lcccccc}
	\tablecaption{Summary of trace ice feature detections \label{tab:ice_dets}}
	\tablecolumns{7} 
	\tablewidth{\textwidth} 
	\tablehead{
        \colhead{Disk} & 
        \colhead{$^{13}$CO$_2$} & 
        \colhead{OCN$^-$} & 
        \colhead{CH$_4$} & 
        \colhead{NH$_3$ / CH$_3$OH} & 
        \colhead{6.8~$\mu$m} & 
        \colhead{14.2~$\mu$m}  
        }
\startdata
HV Tau C & \checkmark & \checkmark & \checkmark & ? & \checkmark & \checkmark \\
Oph 163131 & \checkmark & \checkmark & - & - & \checkmark  & - \\
Flying Saucer & ? & - & - & ? & - & \checkmark \\
ESO-H$\alpha$ 574 & ? & \checkmark & - & - & - & \checkmark \\
LkH$\alpha$ 263C & \checkmark & \checkmark & - & - & \checkmark & \checkmark
\enddata
\tablenotetext{}{\checkmark = Detected, ? = Tentative feature, - = Not detected}
\end{deluxetable}

\subsubsection{9--10 $\mu$m}
Prominent absorptions from NH$_3$ and CH$_3$OH ice fall in the 9--10 $\mu$m range and are commonly detected in protostars \citep{Bottinelli2010}.  None of the disks in our sample exhibit a clear absorption feature in this region.  HV Tau C and Flying Saucer show hints of absorption centered at 9.5 and 9.7 $\mu$m that are compatible with NH$_3$ and CH$_3$OH, respectively.  However, higher signal-to-noise would be needed to determine if these features are real.  Similarly, NH$_3$ and CH$_3$OH were not convincingly seen in the HH 48 NE disk, and their abundances were constrained to be at least $\sim$3$\times$ lower than ISM levels \citep{Sturm2024}.  Because NH$_3$ and CH$_3$OH are among the most common and relatively abundant ice species detected in protostars \citep{Boogert2015}, it is surprising that they are so far elusive in disks.  On the other hand, the disk constraints may be compatible with cometary NH$_3$ and CH$_3$OH abundances, which are typically lower than protostellar values \citep{Mumma2011}.  More work is needed to understand whether there are any observational biases (e.g.~vertical segregation in ice compositions, other radiative transfer effects) that are impeding their detection, or if they are truly under-abundant in disks.

\subsubsection{14.2 $\mu$m}
As noted in Section \ref{subsec:ice_major}, in all disks except Oph 163131, we see a feature around 14.2 $\mu$m neighboring the CO$_2$ bending mode.  We are not aware of any mixture for which CO$_2$ could be sufficiently blue-shifted to explain this feature.  However, other molecules with an OCO bending mode may absorb in this region.  Indeed, HCOOH exhibits an OCO bend centered at 14.2~$\mu$m and with a comparable width to the observed feature \citep{Bisschop2007}, making this a possible candidate.  HCOOH has been identified in protostellar ices at roughly percent levels with respect to H$_2$O via its feature around 8.2 $\mu$m \citep{Rocha2024, Rocha2025}. If confirmed, the detection of HCOOH at 14.2 $\mu$m would imply that the disk solids are enriched in formic acid compared to protostars, where absorption at 14.2 $\mu$m is not evident \citep{Pontoppidan2008}.  This, in turn, could have important implications for the potential to form ammonium salts (i.e.~NH$_4^+$HCOO$^-$) in disk ices, which significantly increase the temperature at which NH$_3$ sublimates and influence the partitioning of nitrogen in planet-forming solids \citep{Bergner2016,Altwegg2020}.  However, detailed modeling is needed to definitively assign the 14.2 $\mu$m feature, accounting for different possible CO$_2$ mixtures, grain properties, and spectral features in other wavelength regions.

\subsubsection{6.0 and 6.85 $\mu$m}
\label{subsec:ice_6um}
\begin{figure}
    \centering
    \includegraphics[width=\linewidth]{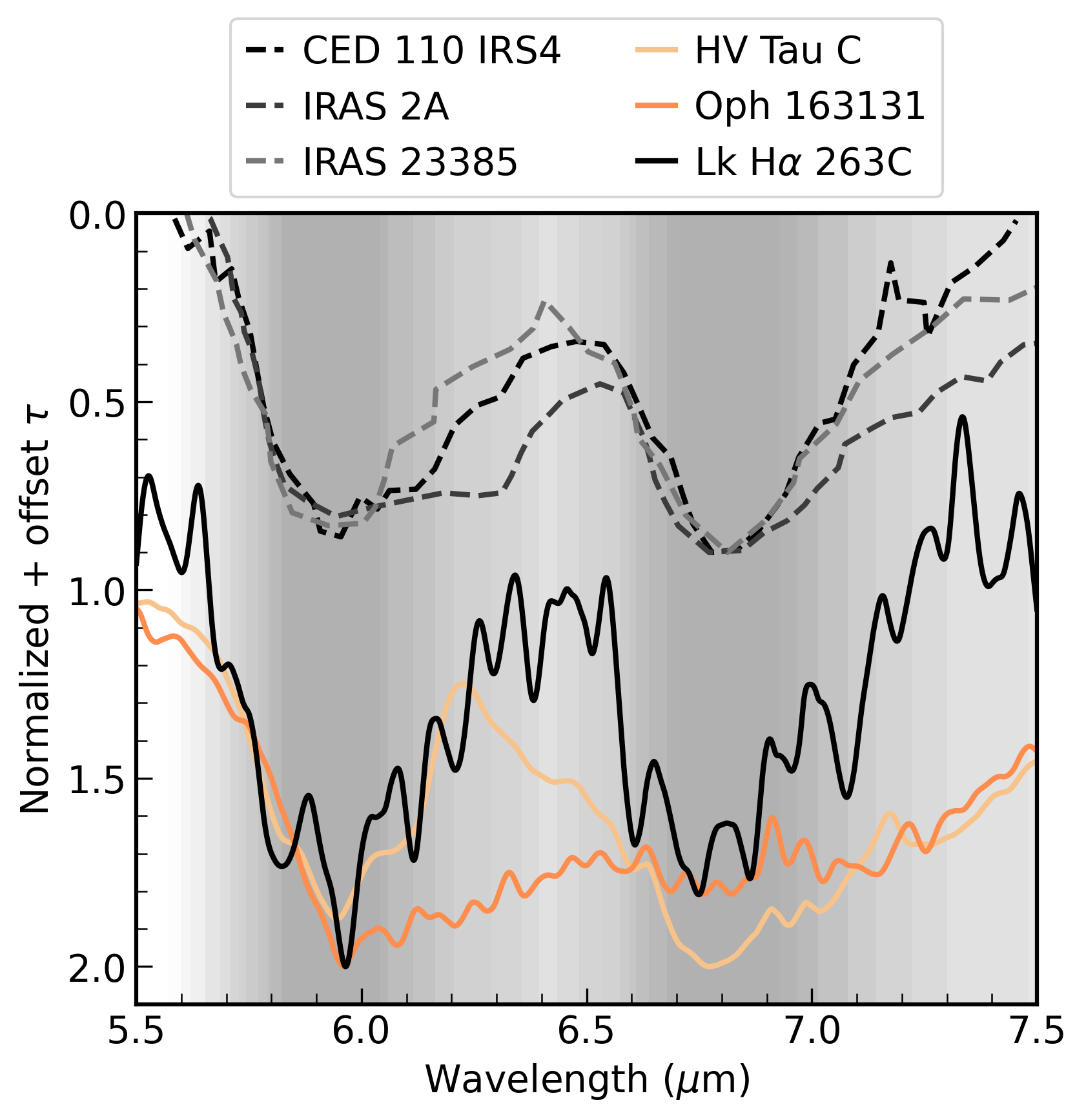}
    \caption{Several disks in our sample (colors) show absorptions in the 6.0 and 6.85~$\mu$m regions similar to those seen along protostellar lines of sight.  For reference, protostar optical depth spectra from \citet{Rocha2024,Rocha2025} are shown with dashed grey lines, and also represented with greyscale intensity.}
    \label{fig:ice_6um}
\end{figure}

The $\sim$5--8 $\mu$m region of protostellar ice spectra is usually dominated by absorptions at 6.0 and 6.85 $\mu$m \citep{Keane2001,Boogert2008}.  Several of the disks in our sample exhibit absorptions at these wavelengths as well, shown in Figure \ref{fig:ice_6um}: HV Tau C, Oph 163131, and Lk H$\alpha$ 263C.  For comparison, we also include recent JWST spectra of low-mass protostars from \citet{Rocha2024} and \citet{Rocha2025}.  In disks, interpretation of this region is somewhat complicated by the possible presence of PAH emission around 6.2 $\mu$m; however, we expect this feature to be small relative to the total depth of the 6.0 and 6.85 $\mu$m absorption features, based on the strength of the 11.3 $\mu$m PAH features seen towards HV Tau C and Oph 163131 (Figure \ref{fig:spec_refbands}).  

In protostars, the H$_2$O bending mode is a contributor to the 6.0 $\mu$m feature, though it cannot fully explain its depth \citep{Gibb2000,Keane2001}.  Water likely contributes to the 6.0 $\mu$m feature in our disk sample as well, given the prominence of its 3 $\mu$m stretching mode in all sources.  Additionally, in protostars, HCOOH/HCOO$^-$ and H$_2$CO are thought to be important contributors to the 6.0 $\mu$m feature, and NH$_4^+$ to the 6.85 $\mu$m feature \citep{Schutte2003, Boogert2008, Rocha2025}.  Given the visual similarity between the protostellar spectra and disk spectra shown in Figure \ref{fig:ice_6um}, it is reasonable to expect that ammonium salts and other organic carriers may also be contributing to the absorptions seen in the disk sample.  

\subsection{Implications for ice compositional evolution}
\label{subsec:interpret}
From our survey of ice features in the JEDIce disks, we can draw several conclusions about the evolution of ice compositions along the journey from protostars to planetesimals.  As in HH 48 NE \citep{Bergner2024}, our observations are incompatible with a pure CO component and favor an apolar CO:CO$_2$ mixture.  This contrasts with protostellar ice compositions, for which the main CO components are pure CO and a polar mixture \citep{Oberg2011b}.  Therefore, we are clearly not probing ices that are pristinely inherited from the protostellar stage.  On the other hand, the presence of mixed ices also excludes a thermally `reset' ice structure, for which a large pure-CO component would be expected due to ice sublimation and recondensation \citep[e.g.][]{Visser2009}.  Rather than a scenario of complete inheritance or reset, we are likely seeing the product of thermal and/or photochemical alteration of the initial protostellar ice reservoir.  

Ice processing in the disk stage could also convert small molecules into more complex species, potentially explaining the non-detection of CH$_4$, NH$_3$, and CH$_3$OH in most of our sample.  Indeed, the conversion of NH$_3$ into ammonium salts is supported by the detection of a strong OCN$^-$ feature towards almost the entire sample and the 6.85~$\mu$m feature towards 3 disks.  These results suggest that disk ices have already evolved towards more comet-like compositions, which are also depleted in CH$_4$, NH$_3$, and CH$_3$OH compared to interstellar ices and are rich in ammonium salts and refractory organics \citep{Mumma2011,Altwegg2020,Fray2016}.  Future quantification of the disk ice compositions will help to reveal the extent of this chemical evolution by the disk stage and the degree to which further processing takes place after the assembly of comets from these icy building blocks.

\section{Summary \& Conclusions}
\label{sec:concl}

This paper presents the near- to mid-IR spectra of five edge-on protoplanetary disks observed as part of the JWST Edge-on Disk Ice (JEDIce) program.  We focus on providing an overview of the complete data set, and carrying out the first comparative study of disk ices viewed by JWST.  Our main findings are as follows.

\begin{itemize}[noitemsep,leftmargin=*]
    \item The spectra show an assortment of gas-phase lines, notably due to molecular hydrogen and CO along with atomic lines from H, Ar, Ne, and Fe.  HV Tau C exhibits a particularly rich gas-phase spectrum, including photochemistry tracers like OH and CH$_3^+$.  
    \item PAH emission features are clearly seen towards ESO-H$\alpha$ 574, HV Tau C, and Oph 163131, expanding the total number of low-mass T Tauri systems with PAH detections to five.  
    \item The peak optical depths for the main H$_2$O, CO$_2$, and CO ice bands vary by factors of $\sim$3--6 across the sample, and similarly the relative depths of different ice bands vary across the sample.  While this suggests that a range of ice abundances and compositions may be present, radiative transfer modeling will be needed to obtain quantitative constraints.
    \item The H$_2$O, CO$_2$, and CO ice band profiles are similar from disk to disk, with subtle variations in the shape and intensity of the band wings.  In all disks, the CO band profile is best explained by an apolar CO:CO$_2$ mixture.  An important implication of the presence of mixed ice is that CO trapping within the CO$_2$ matrix could influence how CO is partitioned between the gas and solid phases.  
    \item In all disks, ice features are detected throughout the entire region where the IR continuum is detected, suggesting that vertically elevated icy grains are the norm in protoplanetary disks.  Across the sample, we see differences in the ice optical depth profiles along the midplane, which remain unexplained.  
    \item The CO optical depth map for Flying Saucer shows a sharp transition in the vertical direction that is not seen for H$_2$O or CO$_2$, and could potentially be a signature of its snow surface.
    \item OCN$^-$ is a common trace ice component with a detection in 4 of the 5 disks.  CH$_4$ is detected towards HV Tau C alone.  There is no clear evidence for NH$_3$ or CH$_3$OH towards any disk, though NH$_3$ hydrates may contribute to the red wing seen on the 3~$\mu$m H$_2$O band.  
    \item Three disks exhibit absorptions around 6.0 and 6.85~$\mu$m similar to those seen in protostars, which are commonly attributed to ammonium salts and other organics like HCOOH and H$_2$CO. We also identify a feature at 14.2~$\mu$m towards 4 disks, which is unlikely to be CO$_2$, but may correspond to the OCO stretch of another molecule like HCOOH.  
    \item The disk ice compositions are neither completely reset nor pristinely inherited from the protostellar stage.  They seem to represent an evolution towards more comet-like ice compositions than protostellar ices.
\end{itemize}

Building on the empirical analysis presented here, future modeling efforts will provide insight into the ice abundances and ice mixtures present in the disks, the gas column densities and excitation conditions, the disk wind and jet properties, and the PAH characteristics and distributions.  Taken together, it is clear that JWST observations of edge-on disks have unique potential to reveal the chemical, physical, and dynamical state of protoplanetary disks, and in turn elucidate the environment in which planetary assembly takes place.

\FloatBarrier
\begin{acknowledgements}

This work is based on observations made with the NASA/ESA/CSA James Webb Space Telescope and the NASA/ESA Hubble Space Telescope. The data were obtained from the Mikulski Archive for Space Telescopes at the Space Telescope Science Institute, which is operated by the Association of Universities for Research in Astronomy, Inc., under NASA contracts NAS 5–26555 and NAS 5-03127.  The specific observations analyzed can be accessed via \dataset[doi:10.17909/hmaf-9z97]{https://doi.org/10.17909/hmaf-9z97}.

Support for Program number 5299 was provided by NASA through a grant from the Space Telescope Science Institute, which is operated by the Association of Universities for Research in Astronomy, Inc., under NASA contract NAS 5-03127. A part of this research was carried out at the Jet Propulsion Laboratory, California Institute of Technology, under a contract with the National Aeronautics and Space Administration (80NM0018D0004).  AN and ED acknowledge support from the Thematic Action `Physique et Chimie du Milieu Interstellaire' (PCMI) of INSU Programme National `Astro', with contributions from CNRS Physique \& CNRS Chimie, CEA, and CNES. DH is supported by the Ministry of Education of Taiwan (Center for Informatics and Computation in Astronomy grant and grant number 110J0353I9) and the National Science and Technology Council, Taiwan (Grant NSTC111-2112-M-007-014-MY3, NSTC113-2639-M-A49-002-ASP, and NSTC113-2112-M-007-027).  JCS was supported by the Heising-Simons Foundation through a 51 Pegasi b Fellowship.

This paper also makes use of data from the ALMA programs 2023.1.00634.S, 2016.1.00771.S, 2023.1.00907.S, and 2022.1.01302.S. ALMA is a partnership of ESO (representing its member states), NSF (USA) and NINS (Japan), together with NRC (Canada), NSTC and ASIAA (Taiwan), and KASI (Republic of Korea), in cooperation with the Republic of Chile. The Joint ALMA Observatory is operated by ESO, AUI/NRAO and NAOJ.
\end{acknowledgements}

\software{
{\fontfamily{qcr}\selectfont Matplotlib} \citep{Hunter2007},
{\fontfamily{qcr}\selectfont NumPy} \citep{Numpy2020},
{\fontfamily{qcr}\selectfont Scipy} \citep{Scipy2020},
{\fontfamily{qcr}\selectfont Astropy} \citep{Astropy2022},
}

\FloatBarrier
\appendix 
\FloatBarrier
Table \ref{tab:calibration} lists the calibration and pipeline details for the observations presented in this work, as well as the elliptical aperture parameters used to extract disk-integrated spectra.  Table \ref{tab:continuum} lists the spectral ranges used to fit global continuum models.

\begin{figure*}
    \centering
    \includegraphics[width=\linewidth]{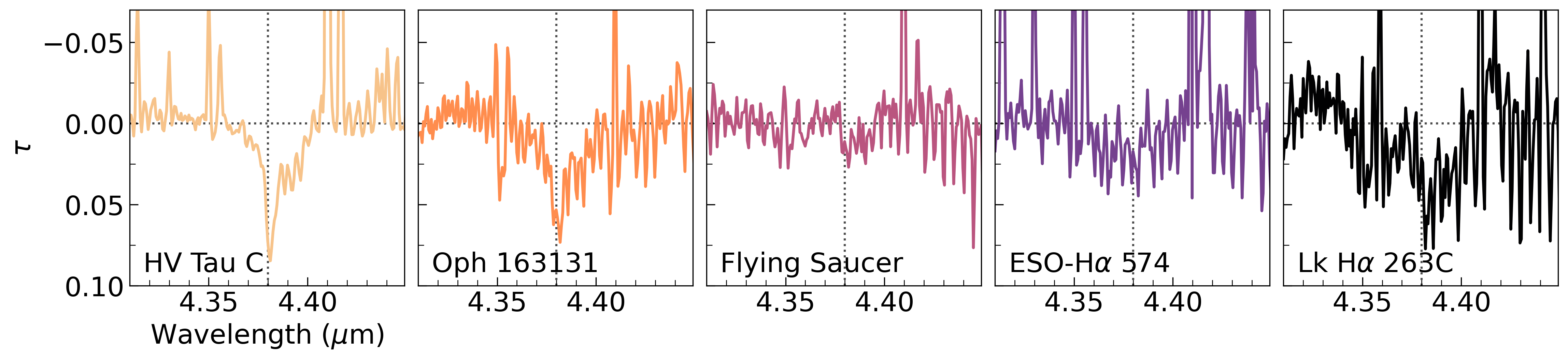}
    \caption{Local optical depth spectra around the $^{13}$CO$_2$ stretching band.  Dotted vertical lines mark the absorption peak at 4.38~$\mu$m.}
    \label{fig:13co2}
\end{figure*}

\begin{figure*}
    \centering
    \includegraphics[width=\linewidth]{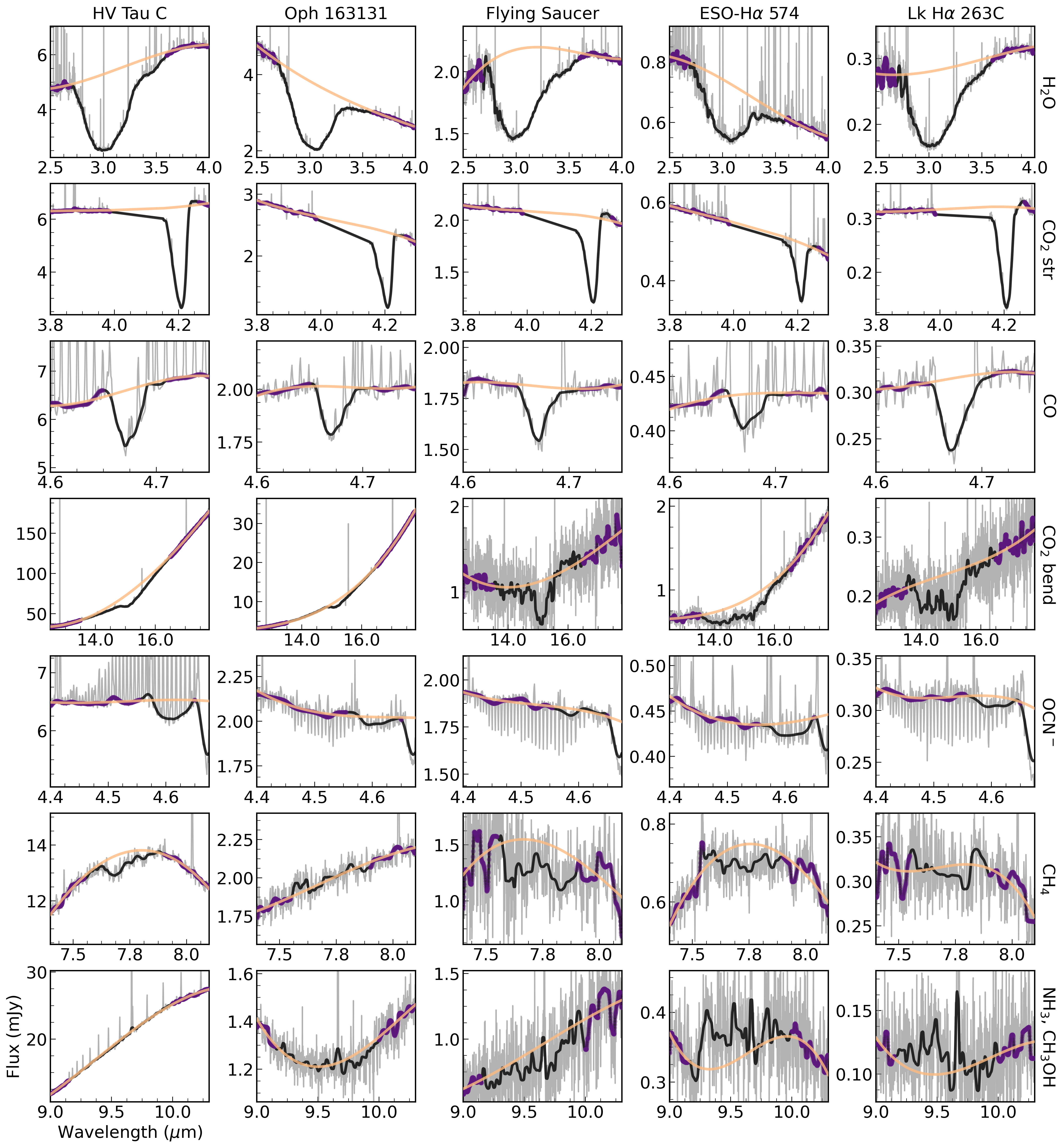}
    \caption{Zoom-in around ice band regions, with grey lines showing the original spectra and black lines showing the smoothed spectra.  Purple points represent the regions used to anchor the continuum, and orange lines show the 3rd-order polynomial continuum models used to produce optical depth spectra.}
    \label{fig:local_cont}
\end{figure*}

Figure \ref{fig:13co2} shows local optical depth spectra around the $^{13}$CO$_2$ stretching band for each disk.  Figure \ref{fig:local_cont} shows zoom-ins of the observed spectra around each ice band discussed in Section \ref{sec:ice}, along with the adopted baseline regions and the continuum models used to produce the local optical depth spectra shown in Figures \ref{fig:ice_major} and \ref{fig:ice_trace}.

\begin{deluxetable*}{lcccccccc}
	\tablecaption{Calibration and spectral extraction details. \label{tab:calibration}}
	\tablecolumns{9} 
	\tablewidth{\textwidth} 
	\tablehead{
        \colhead{Disk}        & 
        \multicolumn{3}{c}{NIRSpec/IFU} & 
        \multicolumn{3}{c}{MIRI/MRS}  & 
        \multicolumn{2}{c}{Elliptical extraction aperture} \\
        \colhead{} & 
        \colhead{Pipeline} & 
        \colhead{CRDS} & 
        \colhead{Obs date} & 
        \colhead{Pipeline} & 
        \colhead{CRDS} & 
        \colhead{Obs date} & 
        \colhead{Major $\times$ minor axis ($''$)} & 
        \colhead{Orientation ($^\circ$) $^a$ }
        }
\startdata
HV Tau C & 1.17.1 & 1322 & 2025-02-10 & 1.18.0 & 1364 & 2023-09-27 & 1.6 $\times$ 1.2 & 29\\
Oph 163131 &1.15.1 & 1293 & 2024-08-18 & 1.17.1 & 1322 & 2025-03-28  & 2.6 $\times$ 1.1 & 137 \\
Flying Saucer & 1.17.1 & 1322 & 2024-08-25 & 1.18.0 & 1364 & 2023-08-28  & 4.0 $\times$ 2.2 & 87 \\
ESO-H$\alpha$ 574 & 1.17.1 & 1322 & 2024-08-23& 1.17.1 & 1322 & 2025-02-18 & 1.4 $\times$ 0.8 & 40 \\
LkH$\alpha$ 263C & 1.17.1 & 1322 & 2024-08-24 & 1.15.1 & 1293 & 2024-09-13 & 1.4 $\times$ 0.8 & 46
\enddata
\tablenotetext{a}{Clockwise from vertical}
\end{deluxetable*}

\begin{deluxetable*}{ll}
	\tablecaption{Global continuum wavelength regions. \label{tab:continuum}}
	\tablecolumns{2} 
	\tablewidth{\textwidth} 
	\tablehead{
        \colhead{Disk}        & 
        \colhead{Continuum wavelengths ($\mu$m)} 
        }
\startdata
HV Tau C &  1.0--2.7, 3.75--3.85, 4.9--5.5, 10.5--11.0, 12.5--13.0, 13.8--14.0, 16.0--23.0, 27.0--27.2 \\
Oph 163131 & 1.0--2.7, 3.75--3.85, 4.9--5.5, 8.0--8.1, 12.5--13.0, 13.8--14.0, 16.0--23.0, 27.0--27.2 \\
Flying Saucer & 1.0--2.7, 3.75--3.85, 4.9--5.5, 7.0--7.1, 11.0-12.0, 13.8--14.0, 16.0--23.0, 27.0--27.2 \\
ESO-H$\alpha$ 574 & 1.0--2.7, 3.75--3.85, 4.9--5.5, 9.0--10.5, 12.5--13.0, 13.8--14.0, 16.0--23.0, 27.0--27.2 \\
LkH$\alpha$ 263C & 1.0--2.7, 3.75--3.85, 4.9--5.5, 7.7--7.9, 13.8--14.0, 16.0--23.0, 27.0--27.2
\enddata
\end{deluxetable*}

\FloatBarrier

\bibliography{references}
\bibliographystyle{aasjournalv7}

\end{document}